\documentclass[a4paper,11pt]{article}
\usepackage[margin=1.14in]{geometry}
\usepackage{tabularx}
\usepackage{multirow}
\usepackage{arydshln}
\usepackage{pdfpages}
\usepackage{float}
\makeatletter \renewcommand{\@dotsep}{10000} \makeatother
\usepackage{wrapfig}
\usepackage{authblk}
\usepackage{lmodern}
\usepackage{amsmath, amssymb, amsthm}
\usepackage{mathtools}
\usepackage{bm}
\usepackage{booktabs}
\usepackage{multirow}
\usepackage{array}
\usepackage{subcaption}
\usepackage{hyperref}
\usepackage{enumitem}
\usepackage{microtype}
\usepackage{parskip}
\usepackage{fancyhdr}
\usepackage{titlesec}
\usepackage{abstract}
\usepackage[colorinlistoftodos, textwidth=2.42cm, size=scriptsize]{todonotes}

\newcommand{\kt}{k_{t}}
\newcommand{\Ldark}{\Lambda_{\rm dark}}
\definecolor{darkred}{rgb}{0.6, 0, 0}
\hypersetup{
	unicode,
	colorlinks,
	breaklinks,
	urlcolor=darkred,
        linkcolor=black,
        citecolor=darkred,
}

\newcommand{\be}{\begin{eqnarray}}
\newcommand{\ee}{\end{eqnarray}}
\def\be{\begin{equation}}
\newcommand{\rinv}{r_{\rm inv}}
\newcommand{\mZp}{m_{Z^\prime}}
\def\ee{\end{equation}}
\def\bea{\begin{eqnarray}}
\def\eea{\end{eqnarray}}

\newcommand{\gsim}{\;\raisebox{-0.9ex}{$\textstyle\stackrel{\textstyle >}{\sim}$}\;}
\newcommand{\lsim}{\;\raisebox{-0.9ex}{$\textstyle\stackrel{\textstyle<}{\sim}$}\;}
\def\lsim{\raise0.3ex\hbox{$\;<$\kern-0.75em\raise-1.1ex\hbox{$\sim\;$}}}
\def\gsim{\raise0.3ex\hbox{$\;>$\kern-0.75em\raise-1.1ex\hbox{$\sim\;$}}}

\usepackage{epsfig}
\usepackage{slashed}
\usepackage[utf8]{inputenc}
\usepackage{changepage}
\usepackage{dcolumn}

\usepackage[dvipsnames]{xcolor}

\usepackage[sort&compress,numbers,square]{natbib}
\theoremstyle{plain}

\theoremstyle{definition}

\usepackage{graphicx}
 
\title{Hunting the Unseen: Deep Learning Analysis \\ for Semi-Visible Jet Tagging}
 
\author[a,b,c]{Miguel A. Avenda\~no-Bernal\footnote{M.A.Avendano-Bernal@soton.ac.uk}}
\author[c]{Srinandan Dasmahapatra\footnote{sd@ecs.soton.ac.uk}}
\author[d]{Ahmed Hammad\footnote{hammad@kias.re.kr}}
\author[a,b,e]{Stefano Moretti\footnote{Stefano.Moretti@cern.ch}}
\author[f,g,h]{Mihoko Nojiri\footnote{mihoko.nojiri@gmail.com}}
\author[i,j]{Michael H. Seymour\footnote{Michael.Seymour@manchester.ac.uk}}
\author[b]{Claire Shepherd-Themistocleous\footnote{Claire.Shepherd@stfc.ac.uk}}

\affil[a]{School of Physics \& Astronomy, University of Southampton,
Southampton, SO17 1BJ, UK}

\affil[b]{Particle Physics Department, Rutherford Appleton Laboratory,\newline
Chilton, Didcot, Oxon, OX11 0QX, UK}

\affil[c]{School of Electronics \& Computer Science, University of Southampton,\newline
Southampton, SO17 1BJ, UK}

\affil[d]{Center of AI and Natural Science, KIAS, Seoul 02455, Korea}

\affil[e]{Department of Physics \& Astronomy, Uppsala University,Uppsala, Box 516, 75120, Sweden}

\affil[f]{Theory Center, IPNS, KEK,  1-1 Oho, Tsukuba, Ibaraki 305-0801, Japan.}
\affil[g]{The Graduate University of Advanced Studies (Sokendai), 1-1 Oho, Tsukuba, Japan.}
\affil[h]{Kavli IPMU (WPI), University of Tokyo, 5-1-5 Kashiwanoha, Kashiwa, Chiba 277-8583, Japan.}

\affil[i]{Department of Physics \& Astronomy, University of Manchester,
Manchester, M13 9PL, UK}
 
\affil[j]{Theoretical Physics Department, CERN, 1211 Geneva 23,
Switzerland}

\date{}
\begin{document}
\maketitle
\vspace{-2ex}

\begin{abstract}
Semi-Visible Jets (SVJs) constitute a distinctive collider signature of strongly interacting dark sectors, embedding Dark Matter candidates, wherein jets contain both visible Standard Model objects and invisible dark hadrons, giving rise to correlated jet activity and missing transverse momentum. In this work, we investigate SVJs produced through a heavy $Z^\prime$ mediator and perform an study over a representative set of benchmark scenarios spanning different mediator masses and dark sector parameters in the context of so-called Hidden Valley Models. To characterise the signal, we combine global event kinematics with jet substructure observables, including the  primary Lund Jet Plane (LJP), the two-point energy correlation, angularity, and charged hadron multiplicity. These representations are used to train five Deep Learning classifiers for SVJ vs standard jet discrimination: a Vision Transformer operating on LJP images, a JetLOV network based on a hierarchical clustering tree, a Multi-Layer Perceptron using high level observables, and two multimodal networks that combine the image-based or hierarchical representations of the radiation pattern with the high jet-level observables. This enables a direct combination of global kinematics, radiation patterns, and jet clustering structure. We find that global kinematic observables outperform the LJP and hierarchical jet representations, with the latter providing stronger discrimination than LJP images. Combining these complementary representations with global kinematics yields the best overall performance.  More broadly, this study shows that unlocking the full discovery potential of SVJs would benefit from going beyond global kinematics to exploit the rich information encoded in their internal structure, providing a benchmark for future searches  at the Large Hadron Collider.

\end{abstract}
\newpage
\noindent\rule{\textwidth}{1pt}
\tableofcontents
\noindent\rule{\textwidth}{0.2pt}
\maketitle \flushbottom
\vspace{4mm}

\section{Introduction}
\label{sec:intro}

The nature of Dark Matter (DM) remains one of the most important open questions in particle physics. Astrophysical and cosmological observations provide strong evidence for its existence, yet its particle content and its interactions with the Standard Model (SM) are still unknown. Collider experiments offer a complementary path to direct and indirect detection searches.  In fact, if DM couples to SM particles, it can be produced, e.g., in proton-proton collisions at the Large Hadron Collider (LHC) and inferred from an imbalance in the visible final state, typically reported as Missing Transverse Energy (${\slashed{E}}_T$ or MET).

Most collider searches for DM have relied on simple so-called mono-$X$ topologies~\cite{Abdallah:2016vcn,Hammad:2022lzo,Costa:2026ucl,Bernreuther:2018nat,Bhattacharya:2022qck}, in which a single SM object recoils against an invisible system (e.g., a  jet~\cite{Barducci:2016fue,Godbole:2016mzr,Roy:2025pht}, photon~\cite{Abdallah:2015uba,Macesanu:2005wj,Roy:2024yoh}, $Z$ or $W$ boson~\cite{Carpenter:2021jbd,Yu:2014ula,Bell:2012rg}, etc.). These searches generally assume that the DM candidate is a single, weakly interacting particle with no further internal structure. More recent theoretical work has broadened this picture by proposing hidden, or dark, sectors with their own rich dynamics \cite{han2008phenomenology,cohen2017lhc,beauchesne2019dark}. A well motivated class of these scenarios introduces a new confining gauge group, often referred to as Hidden Valley Models (HVMs) \cite{cohen2015semivisible,liu2025semi,cazzaniga2024phenomenology,albouy2022theory,carmona2025dark}, which mimics standard Quantum Chromo-Dynamics
(QCD) but is only weakly coupled to the SM through a mediator particle. Such a dark sector can undergo its own parton shower and hadronisation, producing a spectrum of dark hadrons, some of which decay back to visible SM particles.

Among the signatures that emerge from this class of models, SVJs are particularly interesting. In this scenario, a massive resonance, such as a $Z^{\prime}$ boson, decays into a pair of dark quarks, which shower and hadronise within the dark sector before producing a mix of visible SM hadrons and invisible dark hadrons. Crucially, both components end up inside the same jet cone, rather than as separate visible and invisible objects. This makes SVJs distinct from conventional jet  signatures and means that discriminating these from the QCD background might benefit from information not only from the global kinematics of the event but also from the internal structure of the jet itself.

A complication, though, is that the dark sector introduces a large number of poorly constrained parameters, such as the dark hadron masses $m_{{\psi}_d}$, the invisible fraction of dark hadrons $r_{\rm inv}$, and the dark hadronisation scale $\Ldark$. Each of these parameters can reshape the kinematics of the resulting jets in a different way, which makes it difficult to design a single search strategy, or a single Benchmark Point (BP) (over the HVM parameter space), that captures the full range of possible SVJ signatures. A systematic study therefore needs to scan across HVM parameter space, rather than rely on one or more fixed BPs.

Searches for SVJs by the ATLAS and CMS collaborations~\cite{tumasyan2022search,aad2024search,aad2025search,cms2026search} at the LHC 
have already explored parts of this parameter space, largely by using global event and leading-jet observables. Comparatively little attention has so far been given to jet substructure observables~\cite{faucett2022learning}, or to variables that capture observables involving MET \cite{kar2021exploring}, such as the global event MET and the azimuthal angle between the leading jets and the MET, both of which may be sensitive to how the dark sector parameters shape the jet. Alongside this, Machine Learning (ML) analyses started to test these signatures and theoretical developments have looked at self-supervised learning using Transformers \cite{favaro2025semi}, with the addition in recent years of Graph Neural Networks (GNN)~\cite{bernreuther2021casting,bhardwaj2024equivariant}.

In this work, we address this gap with a two component analysis of SVJs produced through a resonant $Z^{\prime}$ mediator, combining global jet kinematics with jet substructure observables. On the global side, we study the transverse momentum, azimuthal angle, and pseudorapidity of the leading
jets, together with the global MET and its azimuthal separation from the leading jets. On the substructure side, we use the Lund Jet Plane (LJP) \cite{Dreyer:2018nbf}, the Energy-Energy Correlation (EEC) function, the angularity, and the charged hadron multiplicity to probe the radiation pattern inside the jet. We generate a
set of 18 SVJ BPs spanning two $Z^{\prime}$ masses and variations of the dark quark mass, the invisible fraction, and the dark hadronisation scale, alongside a QCD background sample, and use these observables to characterise the influence of these dark sector parameters on observable properties of the jets. Building on this, we use the resulting features to train and compare five Deep Learning (DL) classifiers: a Vision Transformer (ViT) \cite{dosovitskiy2020image} with JetLOV \cite{Diaz:2023otq} acting on LJP images, a MultiLayer Perceptron (MLP) acting on high level kinematic and substructure observables, a fusion network combining the ViT, and an MLP and another fusion network with JetLOV and MLP.

The paper is organised as follows. In the next section, we discuss the dynamics of SVJs, then we proceed to describe how we characterise their structure. After which we present our DL analysis and the ensuing results. We finally  discuss the latter and conclude.

\section{Dark Shower Models and SVJs}
\label{sec: Theory}

The HVM~\cite{Strassler:2006im,Han:2007ae} provides a well motivated realisation of hidden sector physics, wherein a dark sector, neutral under the SM gauge interactions, communicates with the visible sector through a heavy mediator. Depending on the ultraviolet completion, the mediator may correspond to a massive gauge boson ($Z^\prime$), an extended Higgs sector, or another heavy messenger field. The hidden sector subsequently undergoes its own dynamics, producing bound states commonly referred to as dark hadrons. Such hidden sectors arise in a broad class of DM models and can give rise to a diverse range of collider signatures, including emerging jets, SVJs, mono-$X$ final states, Higgs portal production and decay, displaced vertices, and soft unclustered energy patterns ~\cite{Cohen:2015toa,Knapen:2016hky}.

Dark showers arise when the hidden sector is  subject to a confining QCD-like gauge interaction, so that the dark sector contains fermions and gauge bosons that play roles analogous to quarks and gluons in standard QCD. As in the SM, the corresponding gauge coupling runs with the energy scale and becomes non-perturbative at the dark confinement scale. Once a pair of dark quarks is produced, it undergoes a parton shower followed by hadronisation, forming dark mesons (and baryons) in close analogy with QCD. Some of these hadrons are protected by an unbroken dark flavour symmetry and are therefore stable: i.e.,  they are likely to be measured at the LHC as MET. The remaining hadrons are unstable and decay back to SM particles, typically producing quarks that subsequently hadronise into ordinary jets. 

\begin{figure}[!ht]
    \centering
    \includegraphics[width=0.75\linewidth]{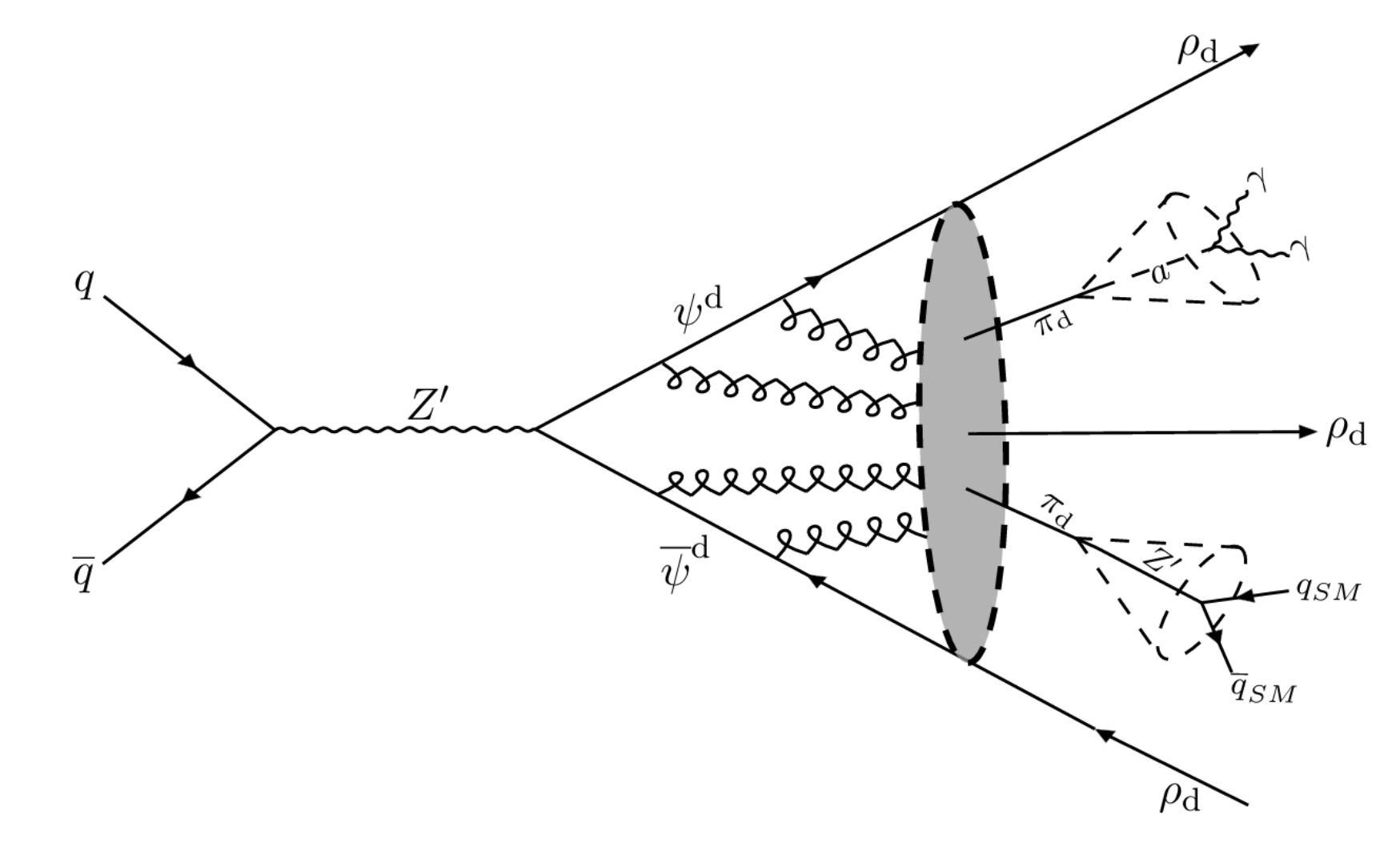}
    \caption{Generic Feynman diagram of a $Z^\prime$ mediated dark shower produced in the $s$-channel for SVJ production. After the decay of the $Z^{ \prime}$, dark quarks are produced forming a QCD-like dark parton shower, then dark hadronisation (represented in the grey area) takes place, ultimately producing both stable and unstable dark hadrons, the latter decaying into SM objects.}
    \label{fig:feynman}
    \vspace{-1ex}
\end{figure}

The collider phenomenology of this type of scenarios is determined by how this hidden sector is connected to the SM. Dark quark pairs can be produced either through a non-resonant $s$-channel QCD process or through the resonant production of the aforementioned mediator. In this work, we focus on the $pp$-collision-induced resonant production channel, where a $Z^\prime$ acts as a portal to the hidden sector and initiates the subsequent parton shower and hadronisation. The collider considered is the LHC. The corresponding Feynman diagram is shown in figure \ref{fig:feynman}. 

The resulting dark hadrons can be either stable or unstable. The unstable hadrons decay back into SM particles, while the stable ones escape the detector and contribute to the MET. The relative abundance of stable dark hadrons is quantified by the invisible fraction, $\rinv$, defined as
\begin{equation}
\rinv \equiv
\left\langle
\frac{n_{\rm dark;hadrons;stable}}
{n_{\rm dark;hadrons;total}}
\right\rangle,
\label{eq}
\end{equation}
where the average is taken over all events. As $\rinv\to0$ the event contains progressively more visible particles and approaches an ordinary QCD di-jet topology; as $\rinv\to1$ the event contains progressively more invisible particles and approaches a  MET  topology.

In the HVM scenario considered here,  the dark  sector contains several free parameters that may alter the kinematic behaviour of the event after the $Z'$ decay, i.e. during the showering and hadronisation stages. We focus on the three parameters  that  have some influence on the internal structure of the SVJ, as follows.
\begin{enumerate}
    \item \textbf{Dark quark mass $m_{\psi_{\rm d}}$}: determines the characteristic mass scale of the dark hadronic states. A variable dark quark mass modifies the kinematics of the shower and hadronisation process, affecting the momentum distribution and multiplicity of the resulting dark hadrons. In particular, heavier stable dark hadrons generally carry a larger fraction of the dark sector energy, leading to enhanced effects in observables sensitive to the visible versus invisible energy balance and MET distribution.   
   \item \textbf{Invisible fraction $r_{\rm inv}$}: controls the fraction of dark hadrons that remain stable and escape detection, thereby determining the balance between visible and invisible components of the dark shower. We expect that increasing(decreasing) $r_{\rm inv}$ will lead to a  larger(smaller) MET and modify the kinematic properties of visible jets. The impact is particularly relevant for SVJ  observables, where the invisible energy/momentum carried by stable dark hadrons can influence the kinematics of the jets obtained from the visible components. For larger $Z^\prime$ masses, the increased available energy should enhance the phase space for additional radiation, making these effects more pronounced. However, the Pythia Monte Carlo simulation that we will be using does not have an explicit $r_{\rm inv}$ parameter. Instead, we adjust the dark hadron branching fractions to obtain the desired value of $r_{\rm inv}$.
   \item \textbf{Dark hadronisation scale $\Ldark$}: sets the characteristic energy scale at which the dark parton shower transitions into dark hadrons. Variations of this parameter modify the hadronisation dynamics, primarily affecting the dark hadron multiplicity, momentum distribution, and the visible versus invisible composition of the final state. However, for jet observables, these effects can  partially be diluted during jet reconstruction, making the sensitivity to $\Ldark$ relatively mild. Observables that retain information about the internal structure of the jet, such as the LJP,  are expected to provide enhanced sensitivity to the details of the hadronisation process. 
\end{enumerate}

To sum up, we expect these parameters to control complementary aspects of the SVJ phenomenology. 
Together, they determine the observable features of dark showers at collider experiments, ranging from missing MET signatures to specific jet substructures. A precise characterisation of these effects requires observables that retain information beyond jet kinematics, motivating the use of jet representations and advanced analysis techniques to identify the subtle patterns associated with SVJ formation.    

\section{SVJ Characterisation}
\label{sec:3}

In this section, we investigate the observable properties of SVJs and identify the features that encode information about the underlying dark sector dynamics. For the modelling of the events, they are generated using \textsc{Pythia8}, with more details of cuts and analysis in section \ref{sec:4.1}, where for most of the jet substructure observables (LJP, two-point correlation function and the angularity), the focus in mainly-based on the leading jet, while the charged hadrons multiplicity will belong to the sum of the leading and the sub-leading jet. Unlike standard QCD jets, SVJs arise from a mixture of visible particles originating from unstable dark hadrons and invisible particles when the latter do not decay. This unique composition leads to characteristic signatures involving both the global event kinematics and the internal structure of the jets. A comprehensive characterisation of SVJs therefore requires observables that probe complementary aspects of the event, from the overall energy balance to the microscopic structure generated during the dark shower and hadronisation stages.

\begin{table}[!ht]
  \centering
  \renewcommand{\arraystretch}{1}
  \resizebox{0.7\textwidth}{!}{%
  \huge
    \begin{tabular}{|l|c|c|c|}
    \hline
    Dataset &  $m_{\psi_v}$ [GeV] & $\Ldark$ [GeV] & $r_{\mathrm{inv}}$ [\%] \\
    \hline
    \color{blue}SVJs\_BP\_01\_mzp\_2.5TeV & \color{red} 10 & 2.50 & 40 \\
    \color{cyan}SVJs\_BP\_02\_mzp\_2.5TeV & \color{red} 50 & 12.50 & 40 \\
    \color{ForestGreen}SVJs\_BP\_03\_mzp\_2.5TeV & \color{red} 100 & 25.0 & 40 \\
    \hline
    \color{magenta}SVJs\_BP\_04\_mzp\_4.5TeV & \color{red} 10  & 2.50 & 40 \\
    \color{brown}SVJs\_BP\_05\_mzp\_4.5TeV & \color{red} 50  & 12.50& 40 \\
    \color{LimeGreen}SVJs\_BP\_06\_mzp\_4.5TeV & \color{red} 100  & 25.0 & 40 \\
    \hline
    \color{olive}SVJs\_BP\_07\_mzp\_2.5TeV & 50 & 20.0 & \color{red} 30  \\
    \color{pink}SVJs\_BP\_08\_mzp\_2.5TeV & 50 & 20.0 & \color{red} 20 \\
    \color{purple} SVJs\_BP\_09\_mzp\_2.5TeV & 50 & 20.0 & \color{red} 50 \\
    \hline
    \color{teal}SVJs\_BP\_10\_mzp\_4.5TeV & 50  & 20.0 & \color{red} 30 \\
    \color{violet}SVJs\_BP\_11\_mzp\_4.5TeV & 50  & 20.0 & \color{red} 20 \\
    \color{Apricot}SVJs\_BP\_12\_mzp\_4.5TeV & 50  & 20.0 & \color{red} 50 \\
    \hline
    \color{DarkOrchid}SVJs\_BP\_13\_mzp\_2.5TeV & 50 & \color{red} 4.0  & 40\\
    \color{Turquoise}SVJs\_BP\_14\_mzp\_2.5TeV & 50 & \color{red} 10.0 & 40 \\
    \color{Bittersweet}SVJs\_BP\_15\_mzp\_2.5TeV & 50 & \color{red} 30.0 & 40 \\
    \hline
    \color{Peach}SVJs\_BP\_16\_mzp\_4.5TeV & 50 & \color{red} 4.0  & 40 \\
    \color{Thistle}SVJs\_BP\_17\_mzp\_4.5TeV & 50 & \color{red} 10.0 & 40 \\
    \textcolor{OliveGreen}{SVJs\_BP\_18\_mzp\_4.5TeV} & 50 & \color{red} 30.0 & 40 \\
    \hline
    SVJs\_BP\_19\_mzp\_3.5TeV &  75 & 2.50 & 60 \\
    SVJs\_BP\_20\_mzp\_3.5TeV & 50 & 12.50 & 65 \\
    SVJs\_BP\_21\_mzp\_3.5TeV & 50 &  20.0 & 40 \\
    \hline
    \end{tabular}
}
\caption{BPs used in the analysis, obtained by varying the three HVM parameters with the largest impact on the observables: the dark quark mass $m_{\psi_{\rm d}}$, the invisible fraction $r_{\rm{\text{inv}}}$, and the dark hadronisation scale $\Ldark$. Variations are indicated in \textcolor{red}{red}. Variation of the $Z^{\prime}$ mass in the range $[2.5,4.5]~\mathrm{TeV}$ is included in the name tag of each BP. The remaining parameter are kept to their default values in the PYTHIA implementation \cite{Bierlich:2022pfr}. BPs 19, 20 and 21 related to the $\mZp~=~3.5~\text{TeV}$ choice will act as testing data since the classifier is trained over a limited range of values for the HVM paramaters and such BP configurations should fall within the used values. The colour scheme used here for the BP name is consistently used in all figures below.} 
\label{tab:SVJs_event_gen_configs}
\end{table}

We first consider global kinematic observables that describe the event topology, including the transverse momentum distribution of the leading jets, MET, and angular correlations between visible and invisible components. These observables are particularly sensitive to the mediator mass, the amount of energy transferred to the invisible sector, and the resulting imbalance between visible and invisible final state particles, respectively. However, global observables alone may not fully capture the complexity of SVJs, as important information about the dark shower evolution can be encoded in the distribution of particles inside the jet.

To access this information, we further investigate jet substructure observables that characterise the internal energy flow and radiation pattern within SVJs (and standard QCD jets). These observables retain sensitivity to the details of the parton shower and hadronisation, and interplay between stable and unstable hadrons in the dark sector. In particular, they provide complementary information to global event level variables and can reveal differences between SVJs and  QCD backgrounds that are not apparent from specific kinematic quantities. The combination of global and local jet information therefore provides a more complete description of SVJ signatures and forms the basis for their discrimination from SM jets in what follows.

To systematically study the dependence of these observables on the parameters of the HVM, we consider a set of 18 BPs generated by varying the aforementioned parameters that have the largest impact on the SVJ phenomenology. The parameters considered include the mediator mass $m_{Z^\prime}$, the dark quark mass $m_{\psi_{\rm d}}$, the invisible fraction $r_{\rm inv}$ and the dark hadronisation scale $\Ldark$. These BPs are chosen to span different regions of the model parameter space and to isolate the individual effects of the dark sector dynamics on the observable properties of SVJs.

Table~\ref{tab:SVJs_event_gen_configs} summarises the parameter space  configurations considered in this analysis. The colour scheme associated with each BP will be used throughout the paper to facilitate the comparison between different dark sector scenarios. The BPs include variations of the mediator mass between $2.5$ and $4.5$ TeV, allowing us to investigate the impact of the available energy scale on the shower evolution and final state observables. For each BP, only the parameters under investigation are modified, while the remaining hadronisation and shower parameters are kept fixed to the default PYTHIA implementation \cite{Bierlich:2022pfr}.

\subsection{Global Kinematic Variables}
\label{subsec: global_variables}

The first step in the characterisation of SVJs is to investigate observables constructed from the global event kinematics. These quantities are directly related to detector level measurements and provide a baseline description of the event topology. For the leading jets, standard kinematic variables such as the transverse momentum $p_T$, energy $E$, pseudorapidity~$\eta$, rapidity $y$, and azimuthal angle $\phi$ provide information about the production mechanism and the available energy scale of the process.

Previous searches by the ATLAS and CMS collaborations~\cite{tumasyan2022search,aad2024search} have explored SVJ signatures using several selections based on these global observables, considering different regions of the HVM  parameter space, including variations of the dark quark mass, mediator mass, and invisible fraction. These studies demonstrated the importance of conventional kinematic variables in constraining exotic signatures. However, SVJs contain additional information associated with the coexistence of visible and invisible dark hadrons, motivating the investigation of observables that explicitly probe this energy imbalance.
In addition to conventional jet kinematics, we focus on MET  and angular correlations between the visible jets and the missing momentum direction. In particular, we consider the azimuthal angle
between the transverse momenta of the leading jets and the MET vector $\slashed{E}_T$. This observable is sensitive to the semi-visible nature of the jets, since the invisible component originates from stable dark hadrons produced within the same dark shower as the visible jet constituents.

The combination of global kinematic variables allows us to study two complementary aspects of the signal. First, it provides insight into how variations of the HVM parameters modify the event topology. Second, it allows us to evaluate their potential discrimination power between SVJs and the QCD background. For this purpose, all 18 BPs introduced in section~\ref{sec:3} are considered, allowing us to explore the dependence of the observables on the HVM parameters.
\begin{figure}[!ht]
    \centering
    \includegraphics[width=0.96\linewidth]{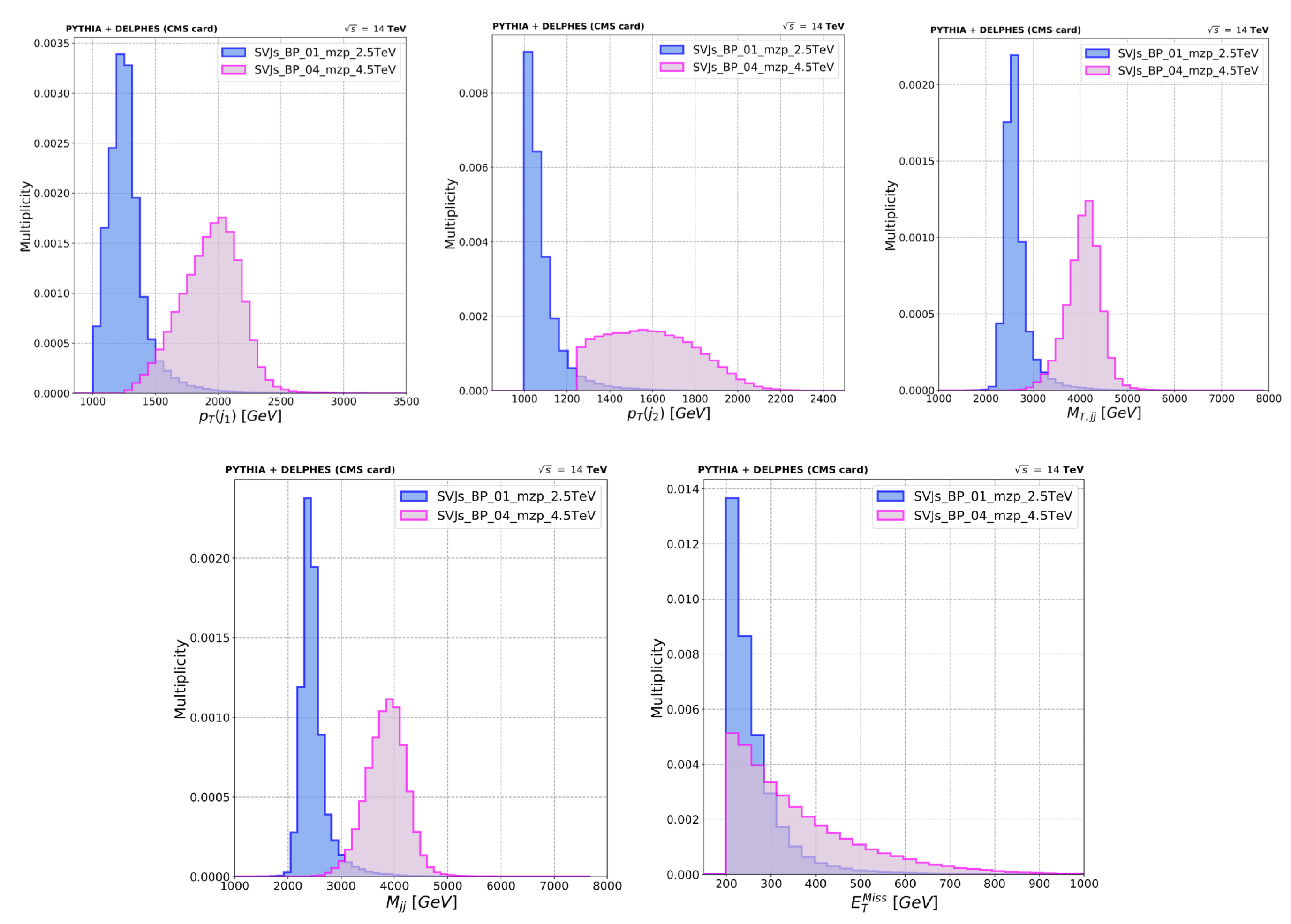}
    \caption{Kinematic variables for two BPs with $Z^\prime$ masses of 2.5~TeV (blue) and 4.5~TeV (red).}
    \label{fig:Global_2500mzp}
\end{figure}
Figure~\ref{fig:Global_2500mzp} shows the distributions of several global observables for BPs with $m_{Z^\prime}=2.5$~TeV and $m_{Z^\prime}=4.5$~TeV. The variables considered include the transverse momentum of the di-jet system, the pseudorapidity distribution, the azimuthal separation between the leading and sub-leading jets, and the invariant mass of the visible di-jet system. The latter is reconstructed from the four momenta of the two visible jets as
\begin{equation}
M_{\rm jj} = \sqrt{ (E_1+E_2)^2- \left| \vec{p}_1+\vec{p}_2 \right|^2 },
\label{Eq:di-jet_mass}
\end{equation}

where $E_{1,2}$ and $\vec{p}_{1,2}$ denote the energies and three-momenta of the two jets, respectively. Since $M_{\rm jj}$ is constructed only from the visible components of the semi-visible jets, it does not include the contribution from stable dark hadrons. Consequently, the reconstructed invariant mass is  shifted away from the true $Z^\prime$ mass scale.

\subsubsection{Transverse Mass } 

To further explore observables sensitive to the invisible component of the event, we consider the transverse mass, $M_{T}$, which is designed for final states containing two visible systems and MET. In the context of SVJs, the two reconstructed jets are treated as two visible branches originating from the decay of the $Z^\prime$ mediator, while the missing transverse momentum arises from stable dark hadrons produced during the two dark showers. The variable for experimental and phenomenological searches is defined as in equation \ref{eq: Transverse_mass},
where $\vec{\slashed{E}}_T$ is the measured missing transverse momentum, the two branches will be measured by taking the square di-jet invariant mass $M_{\rm jj}^2$ shown before, plus the transverse momenta from the di-jet $\vec{{p}}_{T,jj}$, which will contain the visible information from the leading and sub-leading jet transverse momenta defined as $|\vec{{p}}_{T}|~=~|\vec{{p}}_{T,1}~+~\vec{{p}}_{T,2}|$, also shown in figure \ref{fig:Global_2500mzp} for each of the jets.  

The transverse mass of the two branches is then given by
\begin{equation}\label{eq: Transverse_mass}
M_T^2 = M_{\rm jj}^2 + 2 \left(|\vec{\slashed{E}}_{T}|\sqrt{M^{2}_{\rm jj} + |\vec{p}_{T,jj}|^{2}} - \vec{\slashed{E}}_{T}\cdot\vec{p}_{T,jj}\right),
\end{equation}

which is more related to modern approaches on this observable \cite{asadi2026rich}, using the di-jet mass as a first evidence for the smallest parent mass compatible with the observed visible momenta, adding the MET will also consider all the dark sector particles that remain stable. The upper endpoint of the $M_{T}$ distribution will then be closer to the actual parent mass. Therefore, $M_{T}$ provides complementary information to $M_{\rm jj}$ by incorporating the MET associated with stable dark hadrons \cite{cohen2015semivisible}.

In the SVJs scenario considered here, the presence of multiple invisible particles inside each dark shower prevents a direct reconstruction of the invisible system. Nevertheless, $M_{T}$ provides a useful global discriminator by exploiting the correlation between the two visible jets and the missing transverse momentum. As a result, the $M_{T}$ distribution is expected to retain information about the underlying $Z^\prime$ mass scale that is partially lost in the visible invariant mass reconstruction.

\subsubsection{Global MET } 

 The missing transverse momentum is reconstructed from the imbalance of the transverse momentum flow in the event and provides direct sensitivity to the invisible component of the final state. This is described as
\begin{equation}\label{eq: MET}
    \text{MET}~ \equiv~\slashed{E}_{T}~=~\left|\sum \vec{{p}}_{T}\right|~\in~\text{Inv. Particles},
\end{equation} 
where all the invisible particles will belong to the stable dark hadrons from the dark sector or to neutrinos from the SM, merging all the non-electromagnetic-charged particles that could appear on the collision. In QCD multi-jet events, the dominant contribution to $\slashed{E}_{T}$ originates from neutrinos produced in heavy flavour decays and from detector effects. In contrast, for SVJs the missing momentum receives an additional contribution from stable dark hadrons produced during the dark shower and hadronisation processes. These stable states escape the detector and carry away a significant fraction of the energy initially deposited in the dark sector, leading to enhanced high energy tails of MET distributions.

\begin{figure}[!ht]
    \centering
    \includegraphics[width=0.96\linewidth]{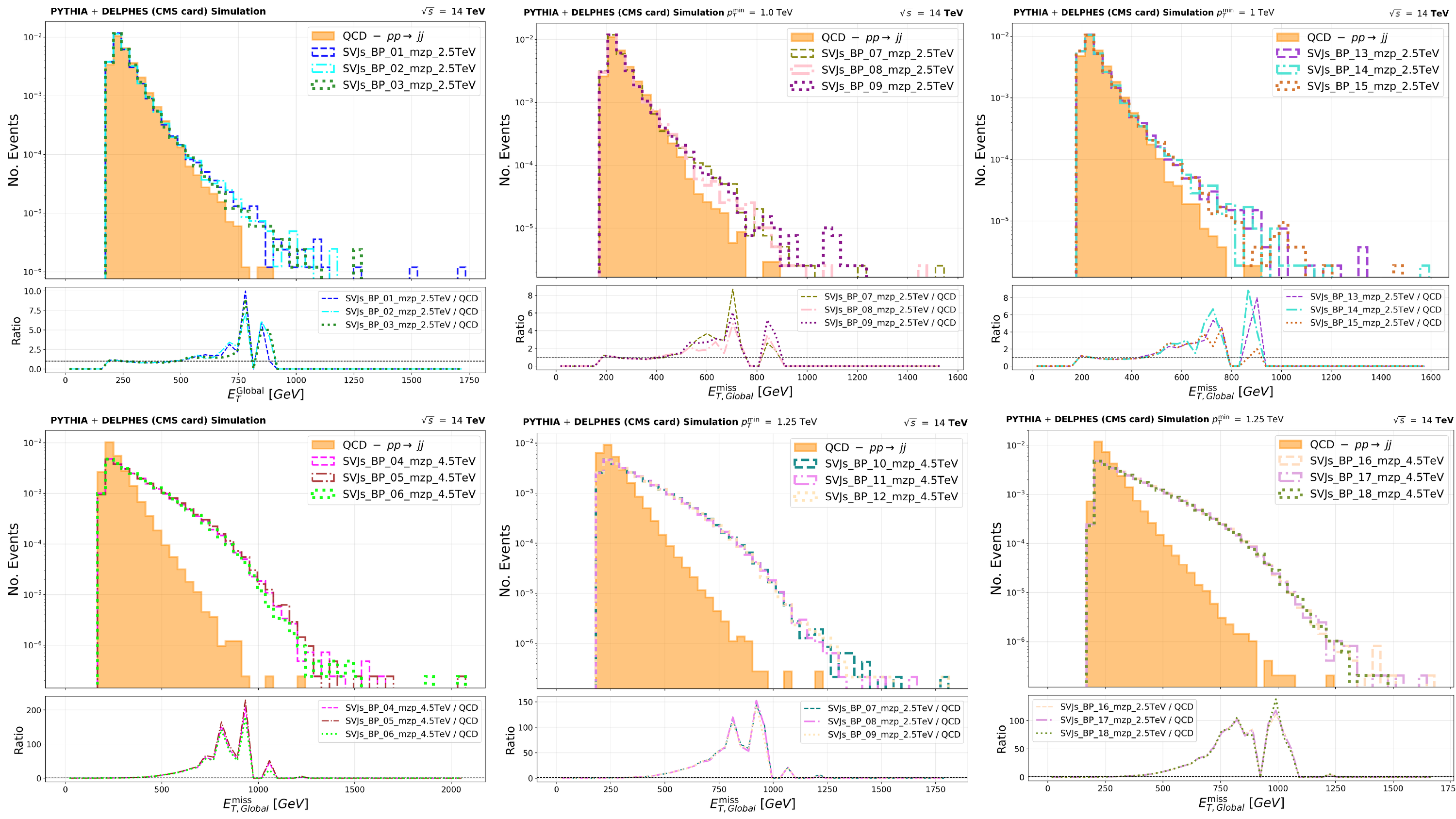}
    \caption{Global MET distribution for QCD jets and SVJs. The colour scheme follows the representation in table \ref{tab:SVJs_event_gen_configs}.}
    \label{fig:MET_studies}
\end{figure}

Figure~\ref{fig:MET_studies} shows the $\slashed{E}_{T}$ distributions for the 18 SVJ BPs compared with the QCD background. The signal samples are separated according to the mediator mass, with $m_{Z^\prime}=2.5$ TeV and $m_{Z^\prime}=4.5$ TeV scenarios exhibiting distinct kinematic behaviours. As expected, increasing the mediator mass leads to a harder energy spectrum and larger MET due to the larger energy available for the production and subsequent evolution of the dark shower. The dependence on the HVM parameters is also visible, with the dark quark mass $m_{\psi_{\rm d}}$ and the invisible fraction $r_{\rm inv}$ producing the largest variations in the $\slashed{E}_{T}$ distribution. The latter directly controls the fraction of dark hadrons that remain stable and therefore determines the amount of energy transferred into the invisible sector.

Although the MET distributions of different SVJ BPs can overlap, they exhibit significant differences with respect to the QCD background. This makes $\slashed{E}_{T}$ a potentially powerful discriminating observable for identifying SVJs signatures. In addition,
the sensitivity of MET searches is expected to become increasingly relevant at higher collider energies. As the available LHC energy increases, heavier dark sector states can be produced, providing additional phase space for dark shower evolution and increasing the potential contribution of stable dark hadrons to the final state momentum imbalance.

\subsubsection{Angular Correlation}

The azimuthal correlation between the missing transverse momentum and the leading jet provides an additional probe of the event topology. We define the complete form of the angular separation as

\begin{equation}\label{eq: diff_azimuthal}
\Delta\phi = \left| \phi_{\rm MET} - \phi_{j_{\rm lead}} \right|,
\end{equation}

where $\phi_{\rm MET}$ and $\phi_{j_{\rm lead}}$ denote the azimuthal angles of the global missing transverse momentum and the leading jet, respectively. The variable measures the directional correlation between the visible and invisible components of the event.

In  QCD multi-jet events, MET is typically associated with neutrinos from heavy flavour decays or detector effects, and therefore tends to exhibit characteristic angular correlations with the reconstructed jets. In contrast, SVJs contain stable dark hadrons produced inside the dark shower, resulting in MET  that is correlated with the visible jet structure. The resulting $\Delta\phi$ distribution therefore provides sensitivity to the semi-visible nature of the signal and is potentially a variable discriminating between SVJs and QCD backgrounds.

\begin{figure}[!ht]
\centering
\includegraphics[width=0.96\linewidth]{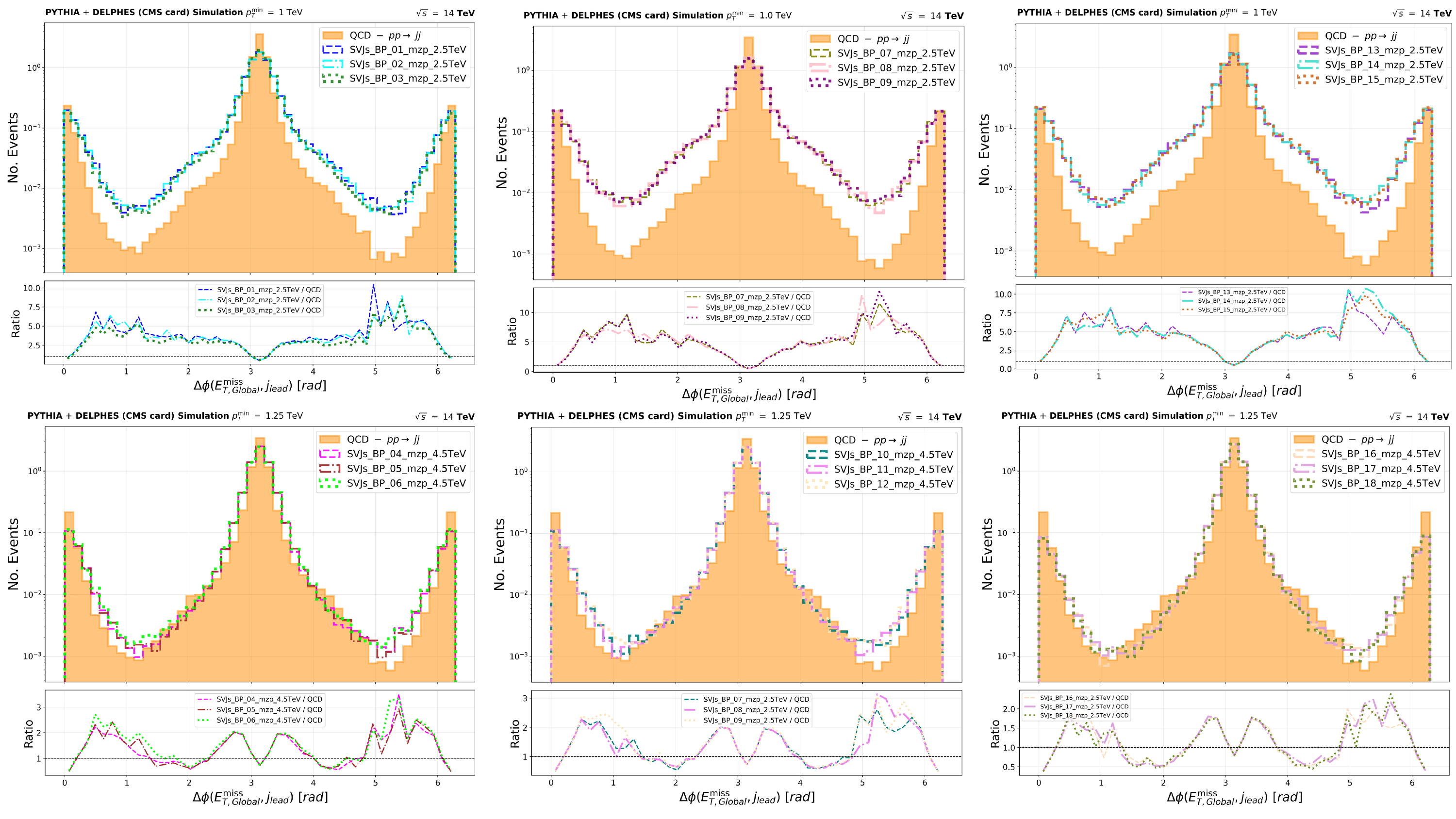}
\caption{Azimuthal separation between the leading jet and the global MET. The colour scheme follows the representation in table \ref{tab:SVJs_event_gen_configs}.}
\label{fig: diff_azimut_cut_ver}
\end{figure}

Figure~\ref{fig: diff_azimut_cut_ver} shows the $\Delta\phi$ distributions for all BPs  in table~\ref{tab:SVJs_event_gen_configs}. 
The dependence on the HVM parameters is mainly driven by the mediator mass, the dark quark mass, and the invisible fraction.
Larger mediator masses provide additional phase space for the dark shower evolution, modifying the angular structure of the final state.

For SVJs identification, $\Delta\phi$ variables are particularly valuable because they retain information about the correlation between visible and invisible components. Unlike MET variables, where events with missing momentum aligned with jets are often suppressed to reduce backgrounds from jet mismeasurement, SVJ scenarios can naturally produce such configurations due to the presence of invisible particles inside the jet cone. 
The distributions shown in figure~\ref{fig: diff_azimut_cut_ver} demonstrate that the considered SVJ BPs can be distinguished from the QCD background using angular information alone. This makes $\Delta\phi$ a useful global observable for classification studies and motivates its combination with additional jet substructure observables that probe the internal structure of the SVJ.

Further studies could investigate the impact of QCD Initial State Radiation (ISR) on all these kinematic variables. Detailed investigation of  ISR contributions would allow a more direct study of the correlation between the dark parton shower, dark hadronisation, and the resulting visible and invisible jet components. Such an investigation is beyond the scope of the present analysis but could provide additional insight into the robustness of our SVJ observables against additional QCD radiation.

\subsection{Substructure Properties of SVJs}

Jet substructure provides information complementary to event level observables based on MET. The latter probe the invisible component directly, but their sensitivity depends on the event recoil topology and decreases for small $\rinv$. Substructure observables instead probe the radiation pattern within the jet and can retain sensitivity when the missing momentum signature is weak. 
Combining complementary observables can therefore provide sensitivity to these effects beyond that of any individual observable.
In this section, we consider four observables: the primary LJP, which resolves the declustering sequence and provides a two dimensional representation of the radiation pattern; angularities, which weight constituents according to their angular distance from the jet axis; the two-point energy correlation function, an Infra-Red (IR)-safe observable sensitive to the angular distribution of energy between pairs of constituents; and the charged hadron multiplicity, which measures the visible hadronic activity after dark hadronisation and decay. These observables are evaluated for the $18$ SVJ BPs listed in table~\ref{tab:SVJs_event_gen_configs} and for a common QCD sample. 

\subsubsection{LJP}
The LJP  provides a two dimensional representation of the radiation pattern inside a jet. It is constructed from the iterative Cambridge/Aachen (C/A) declustering sequence \cite{Dokshitzer:1997in,Wobisch:1998wt}, with each splitting mapped onto a plane spanned by the angular scale and transverse momentum of the softer branch. For a declustering of a pseudojet $j$ into two subjets $j_1$ and $j_2$, with $p_T^{(1)}\geq p_T^{(2)}$, the transverse momentum of the softer emission relative to the harder branch is defined as
\begin{equation}
\Delta R_{12} = \sqrt{(\Delta\eta_{12})^2+(\Delta\phi_{12})^2} \,, \qquad k_t = p_T^{(2)} \Delta R_{12} .
\end{equation}

The corresponding Lund coordinates are
\begin{equation}
    \kappa = \ln\!\left(\frac{k_t}{\mathrm{GeV}}\right),
    \qquad
    \lambda = \ln\!\left(\frac{R}{\Delta R_{12}}\right) \,,
\end{equation}

where $R=0.8$ is the clustering radius of the fat jet. Large values of $\lambda$ correspond to collinear splittings, while smaller values probe wider angle radiation. The coordinate $\kappa$ characterises the hardness of the splitting. The primary LJP is obtained by following the harder branch at each declustering step, resulting in a sequence of points ${(\lambda_k,\kappa_k)}$.

In the soft collinear limit, the perturbative QCD emission density in these logarithmic coordinates takes the approximate form
\begin{equation}
\frac{\mathrm{d}^2 P}
{\mathrm{d}\kappa\,\mathrm{d}\lambda}
\simeq
\frac{\alpha_s(\kt)}{\pi}C_i,
\end{equation}
where $C_i=C_F$ and $C_A$ are the Casimir factors for quark- and gluon-initiated jets, respectively~\cite{Dreyer:2018nbf,Lifson:2020gua}. This relation provides the leading soft collinear baseline for the LJP density of the QCD events. In the fixed coupling approximation, the emission density is uniform in the logarithmic Lund coordinates, resulting in an approximately uniform horizontal band in the reconstructed LJP images for the QCD events. The LJP therefore provides a direct probe of the radiation and hadronisation dynamics of the QCD jet.

This representation is particularly useful for SVJs, where part of the energy generated by the dark sector shower is not reconstructed by the detector. In the HVM considered here, dark quarks produced through the $Z^\prime$ mediator undergo a dark parton shower and subsequently hadronise into dark hadrons. A fraction $\rinv$ of the dark hadrons is stable and invisible, while the remaining component decays to SM particles. Consequently, the reconstructed jet contains only the visible component of the dark shower.
The invisible component modifies the reconstructed LJP in two related ways. First, invisible particles reduce the transverse momentum carried by the visible branches of the shower, shifting the reconstructed splitting scale towards smaller $\kt$. Second, the missing momentum can induce a recoil of the visible jet axis, modifying the reconstructed angular separation and therefore the $\lambda$ coordinate. These effects alter both the density and distribution of splittings in the LJP, providing information that is complementary to the alignment between the jet momentum and the missing transverse momentum.

\begin{figure}[!ht]
\centering
\includegraphics[width=1.\linewidth]{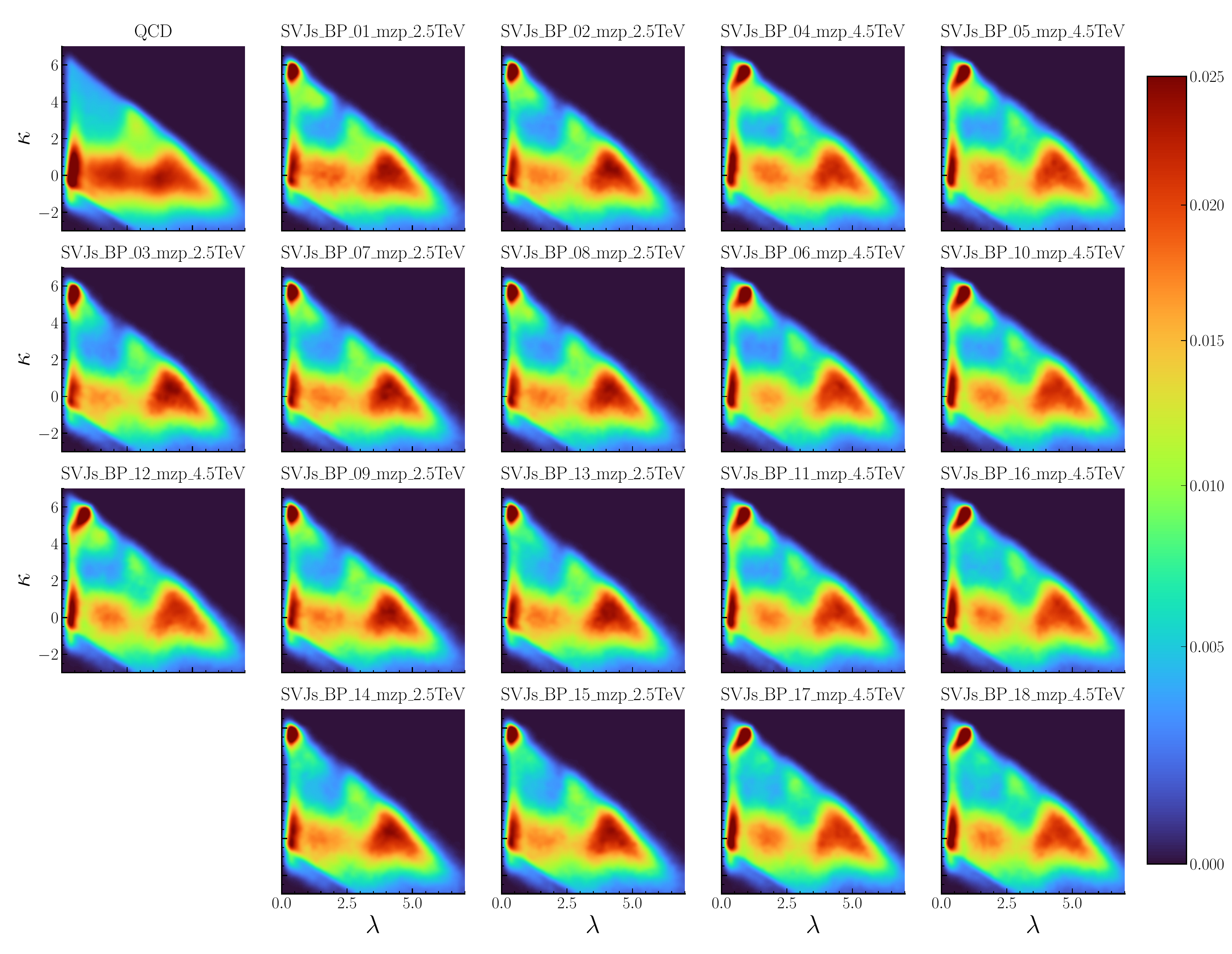}
\caption{Accumulated average primary LJP distributions for 30,000 SVJ events versus the same number of QCD events.
}
\label{fig:lund}
\end{figure}

The dark confinement scale $\Ldark$ provides an additional source of variation in the LJP structure. It sets the characteristic scale at which the perturbative dark shower terminates and non-perturbative dark hadronisation becomes important. For a fixed hard process scale, increasing $\Ldark$ therefore reduces the range of transverse momentum scales over which the perturbative dark shower can develop. Consequently, a larger fraction of the LJP becomes sensitive to non-perturbative effects, modifying the distribution and multiplicity of resolved declusterings \cite{Cohen:2023mya}.

Figure~\ref{fig:lund} shows the LJP distributions obtained for the SVJ samples for each BP. The corresponding QCD distribution provides a reference for identifying modifications induced by the dark shower via the invisible component. A characteristic feature of the SVJ samples is the pronounced dense region, in the top left corner of  the reconstructed LJP, that is consistently present in all different SVJ BPs. This structure is associated with the visible hadrons produced during the dark quark shower and subsequent hadronisation. Due to the relatively large dark quark mass, these hadrons can carry a substantial transverse momentum and are therefore reconstructed as a concentrated population of splittings at large $\kt$. This characteristic structure provides a clear distinction between the internal radiation patterns of QCD jets and SVJs, making the LJP distributions a useful input for the DL analysis.

\subsubsection{SVJ Angularity}

We next consider the angularity as a jet substructure observable. For the reconstructed jets, we use the approximate form \cite{kulkarni2024dark}
\begin{equation}
\tau_{\alpha,\beta} \simeq \sum_{j\in\text{Leading jet}}~\sum_{i\in j} \theta_i^{\alpha} \left( \frac{E_{i}}{E_{j}} \right)^{\beta},
\label{eq: angularity}
\end{equation}
where $\theta_i$ denotes the angular separation between constituent $i$ and the jet axis. In our analysis, we fix $\beta=1$ and vary the angular exponent $\alpha$. We consider three representative values:
\begin{equation}
\alpha\in{0.1,0.8,1.5},
\end{equation}
chosen to probe different regions of the jet radiation pattern and to assess the dependence of the classification performance upon the angular weighting.
Since the analysis is performed using reconstructed hadronic jets, we use the transverse momentum of the constituents and  jet,
$p_{T_i}$ and $p_{T_j}$, respectively. The angular distance is normalised to the jet radius, such that
\begin{equation}
\theta_i =
\frac{\Delta R_{i,j}}{R},
\end{equation}
where $\Delta R_{i,j}$ is the distance between constituent $i$ and the axis of jet $j$. With $\beta=1$, the observable is therefore sensitive to the distribution of the transverse momentum inside the jet, with the value of $\alpha$ controlling the relative contribution of constituents at different angular scales.

\begin{figure}[!ht]
    \centering
    \includegraphics[width=0.96\linewidth]{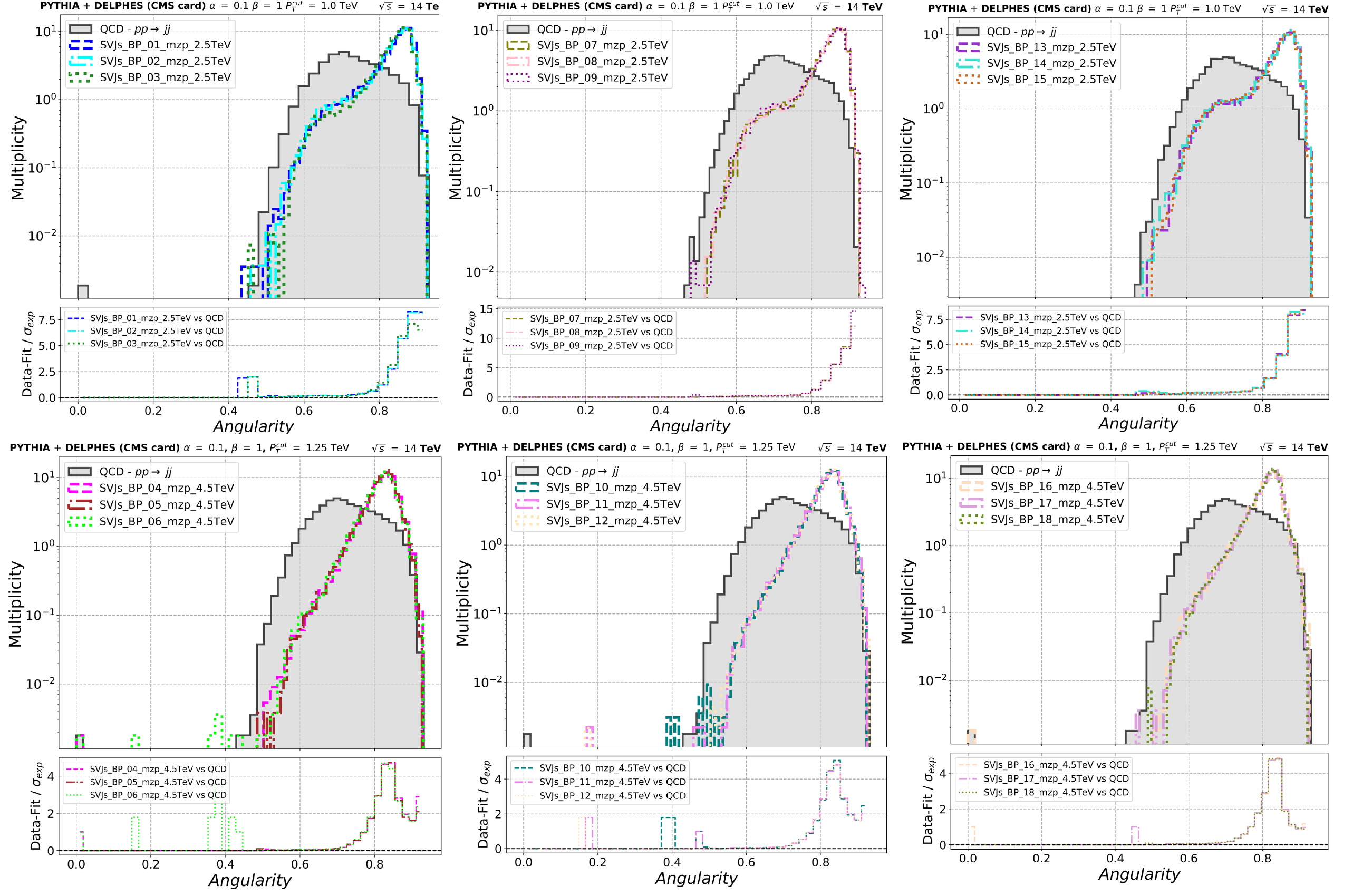}
    \caption{Angularity distribution with $\alpha=0.1$ for the QCD background and the full set of SVJ BPs.}
    \label{fig:Angularity_01}
\end{figure}

Small values of $\alpha$ enhance the contribution from radiation at shorter angular distances, while larger values give greater weight to radiation furthest from the jet axis. Different values of $\alpha$ measures how energy fractions of jet constituents are distributed at different angular distances from the jet axis.  This provides a complementary view of the differences between QCD jets and the dark sector jets considered in this analysis.

We evaluate the angularity distributions for the 18 SVJ BPs listed in Table \ref{tab:SVJs_event_gen_configs}. The same colour scheme is used for the different BPs throughout the analysis, while the QCD background is shown in grey.

Varying $\alpha$ also changes the relative sensitivity of the observables to perturbative and non-perturbative radiation. Considering larger $\alpha$ values, emissions at relative large angular separations will have larger relative weight of emissions, making the observable more sensitive to broad radiation of the jet. Decreasing $\alpha$ will make the observable more sensitive to radiation close to the jet axis, therefore probes more collinear configurations. The three values used here consequently provide complementary information on the internal structure of the jets.

\begin{figure}[!ht]
    \centering
    \includegraphics[width=0.96\linewidth]{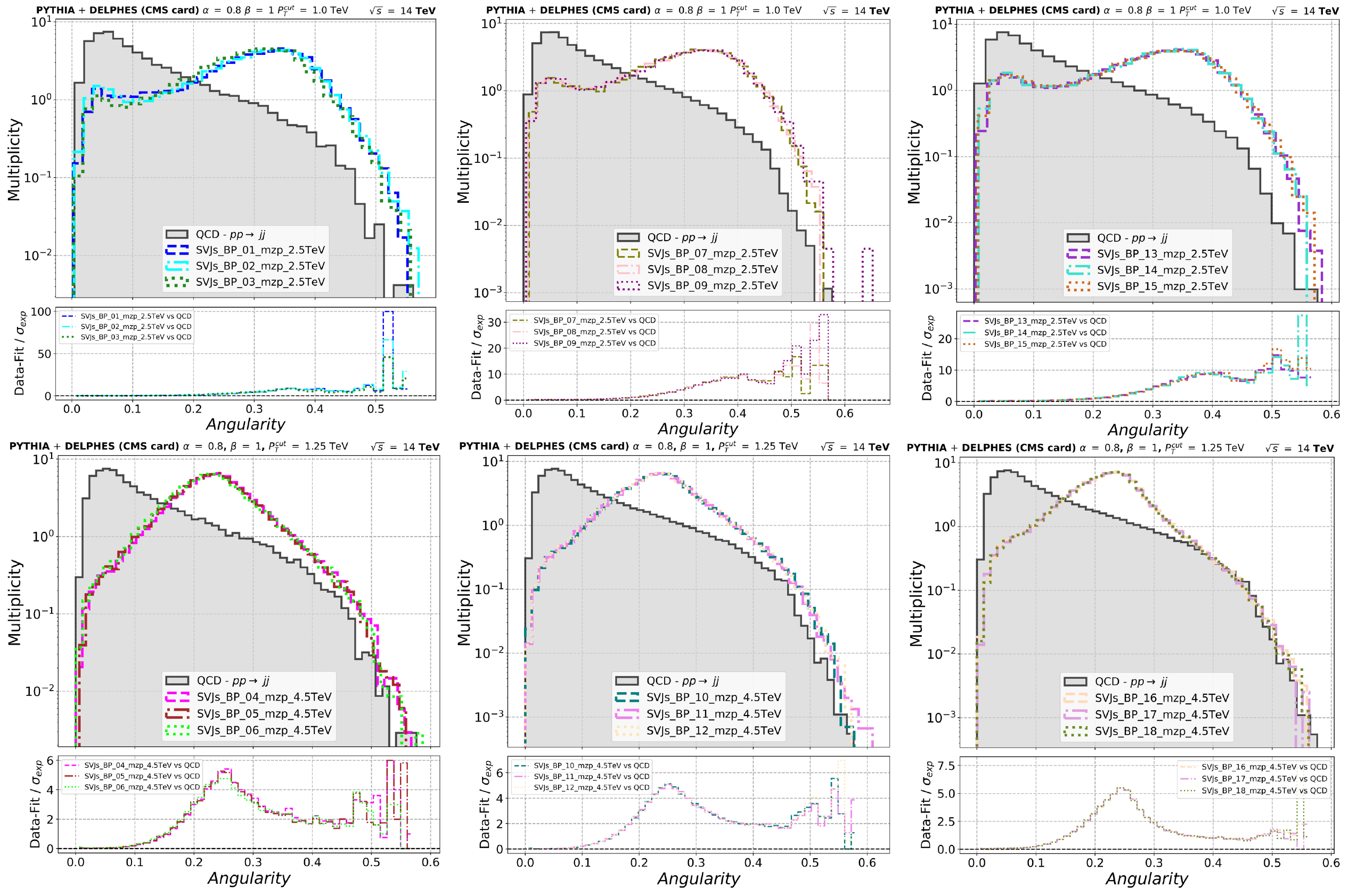}
    \caption{Angularity distribution with $\alpha=0.8$ for the QCD background and the full set of SVJ BPs.}
    \label{fig:Angularity_08}
\end{figure}

\begin{figure}[!ht]
    \centering
    \includegraphics[width=0.96\linewidth]{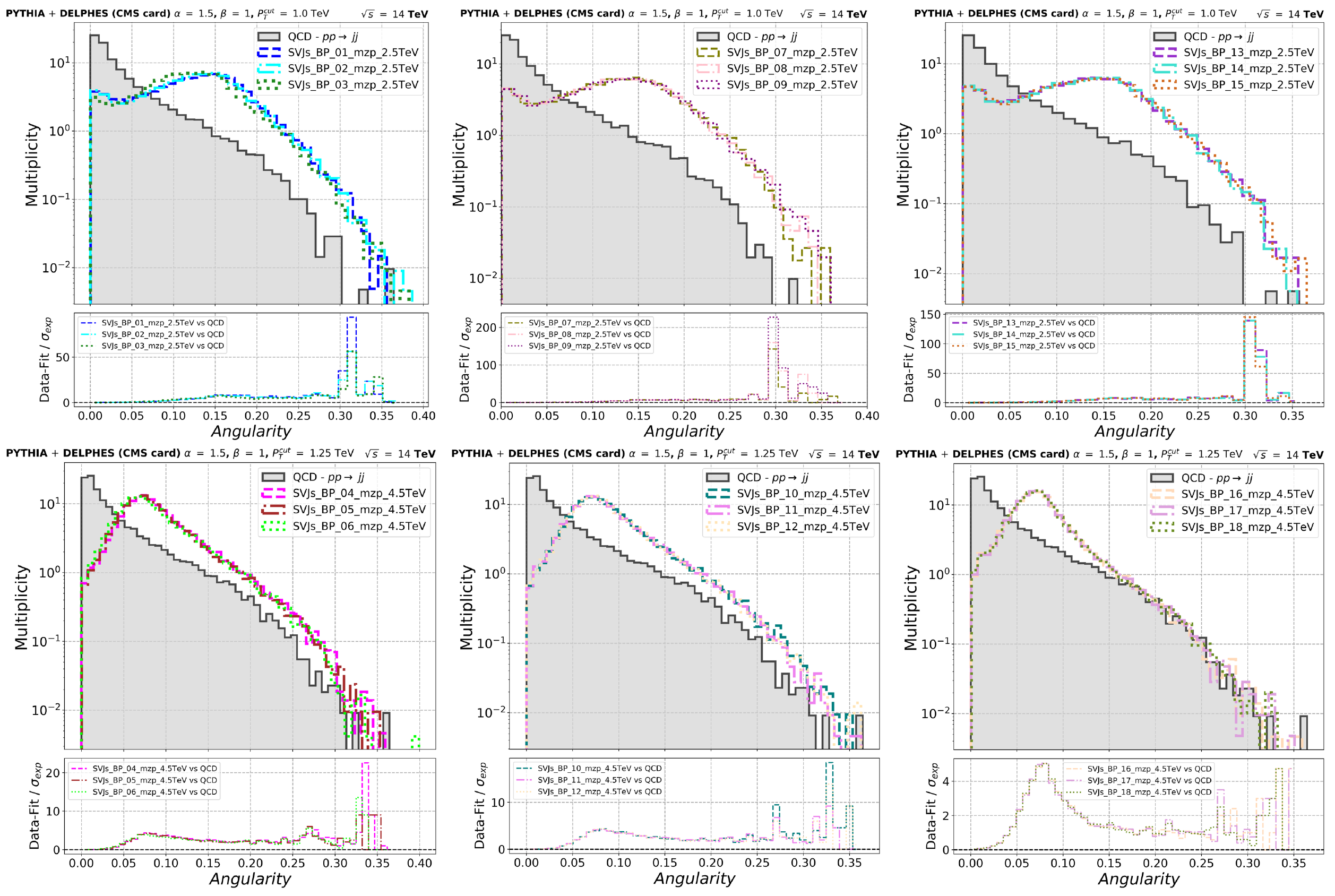}
    \caption{Angularity distribution with $\alpha=1.5$ for the QCD background and the full set of SVJ BPs.}
    \label{fig:Angularity_15}
\end{figure}

For $\alpha=0.1$, radiation close to the jet axis is more strongly weighted, making the observable more sensitive to the collinear structure and to modifications of the jet produced by the dark shower and hadronisation, as shown in figure \ref{fig:Angularity_01}. At $\alpha=0.8$, the weighting becomes more balanced between collinear and wide-angle radiation, providing sensitivity to both the perturbative shower and the onset of non-perturbative effects, as shown in figure \ref{fig:Angularity_08}. For $\alpha=1.5$, the angularity receives substantial contributions from constituents over the wide-angle radiation pattern of the jet, SVJ will exhibit a different behaviour as shown in figure \ref{fig:Angularity_15} because of the modifications at the dark partonic and dark hadronisation level.

In addition to the hard shower and hadronisation dynamics, experimental effects such as pile-up can contribute to the angularity, particularly through soft radiation at large angular separations. Although pile-up is not included in the simulated samples used in this study, these effects should be considered when applying angularity-based observables to experimental data. Appropriate pile-up mitigation and jet-cleaning procedures would therefore be required to preserve the sensitivity to the hard SVJ signal.

\subsubsection{Two-Point Energy Correlation Function}
\label{sec:e2_eec}

\begin{figure}[!ht]
    \centering
    \includegraphics[width=0.96\linewidth]{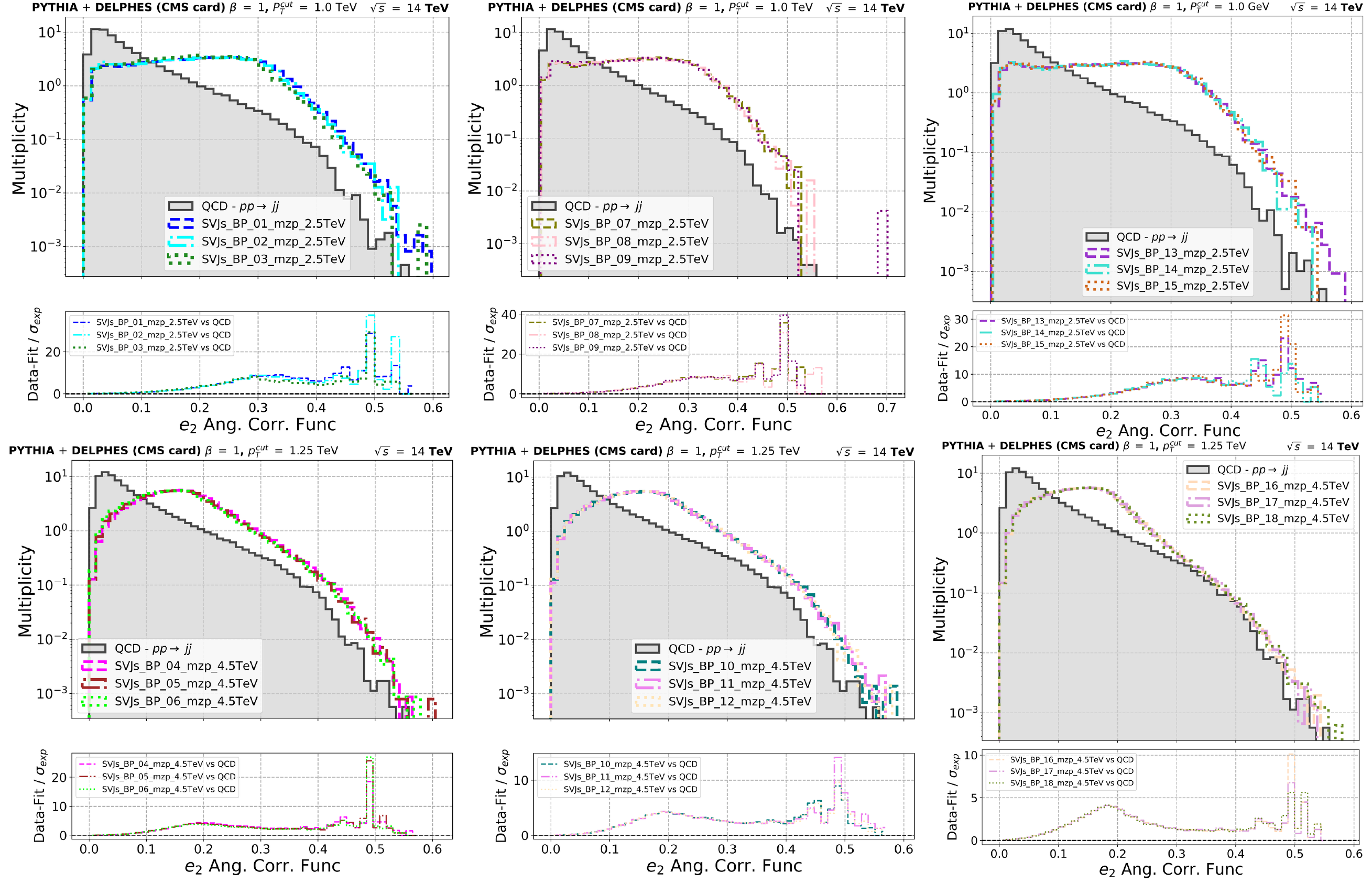}
    \caption{The two-point energy correlation function for the QCD background and the full set of
SVJ BPs. }
    \label{fig:EEC}
\end{figure}

Another jet substructure observable considered in this analysis is the two-point energy correlation function. It probes the angular distribution of the energy inside a jet by correlating pairs of its constituents. We define the two-point correlator as \cite{larkoski2013energy,larkoski2020jet}
\begin{equation}
e_2^{(\beta)}
=
\sum_{i<j\in \mathrm{jet}}
z_i z_j,\theta_{ij}^{\beta},
\label{eq:e2_eec}
\end{equation}
where
\begin{equation}
z_i = \frac{p_{T_i}}{p_{T_\mathrm{jet}}},
\qquad
\theta_{ij} = \frac{\Delta R_{ij}}{R},
\end{equation}
with $\Delta R_{ij}$ denoting the angular separation between constituents $i$ and $j$, and $R$ the jet radius. The energy weighting makes the observable sensitive to the distribution of energy among the jet constituents and provides information on both the perturbative shower and the subsequent hadronisation.
In this analysis, we fix the angular exponent to
\begin{equation}
\beta=1.
\end{equation}
Figure~\ref{fig:EEC} shows the resulting distributions for the 18 SVJ BPs and the QCD background. The same QCD sample is used in all distributions, while the SVJ BPs follow the colour scheme defined in table~\ref{tab:SVJs_event_gen_configs}.
The two-point correlator provides sensitivity to the angular structure of the jet. Small angular separations probe the collinear region of the shower and become sensitive to the transition from perturbative radiation to non-perturbative hadronisation. Larger angular separations probe wider angle radiation and therefore provide information on the structure of the perturbative parton shower. The relative contribution of these regions is controlled by the exponent $\beta$. In the present analysis, $\beta=1$ provides sensitivity to the perturbative radiation pattern while retaining information on the transition towards the non-perturbative regime.

The perturbative and non-perturbative regions can be characterised by comparing the angular scale with the relevant confinement scale. For QCD jets, perturbative radiation is expected when
\begin{equation}
\theta_{ij} \gg
\frac{\Lambda_{\mathrm{QCD}}}{p_{T,\mathrm{jet}}},
\end{equation}
while, for the dark sector, the corresponding condition involves the dark confinement scale,
\begin{equation}
\theta_{ij} \gg
\frac{\Ldark}{p_{T,\mathrm{jet}}}.
\end{equation}
Thus, the two-point correlator can probe both the perturbative dark shower and modifications associated with dark sector hadronisation.

The sensitivity of the correlator to the angular structure can also be understood from its perturbative scaling. In the small angle region, the distribution is governed by the collinear splitting dynamics and receives corrections from the running coupling and anomalous dimensions. Schematically, these effects modify the angular dependence of the correlator according to
\begin{equation}
\frac{\mathrm{d}\sigma}{\mathrm{d}\theta}
\propto
\frac{\alpha_s(\theta)}{\theta^{1-\gamma}},
\end{equation}
where $\gamma$ denotes the relevant anomalous dimension. The precise scaling depends on the perturbative order and on the definition of the correlator. We therefore use the two-point correlator primarily as a complementary probe of the angular radiation pattern rather than assigning a fixed power law behaviour to the full distribution.

An important property of $e_2^{(\beta)}$ is its IR (i.e., soft and collinear) safety for $\beta>0$. Soft emissions are suppressed by the energy weight $z_i z_j$, while collinear splittings do not introduce singular changes in the observable. This makes the correlator theoretically well defined and suitable for comparing the perturbative radiation patterns of QCD and SVJs.

The resulting distributions in figure \ref{fig:EEC}  show characteristic differences between the SVJ BPs and the QCD background. These differences reflect the modified radiation pattern of the dark shower, together with the effects of dark sector hadronisation and the invisible component. As shown previously for the LJP and angularity observables, the dependence of the distribution on the mediator mass and the dark sector parameters provides additional information for distinguishing SVJs from QCD ones. 

\subsubsection{Charged Hadron Multiplicity}
\label{sec:charged_multiplicity}

The charged hadron multiplicity provides a simple measure of the amount of visible hadronic activity inside a jet. We define the observable as the average number of charged hadrons reconstructed within the two leading jets,
\begin{equation}
\langle n_{\mathrm{ch}}\rangle
=
\frac{1}{N_{\mathrm{jets}}}
\sum_{j=1}^{N_{\mathrm{jets}}}
n_{\mathrm{ch}}^{(j)},   \qquad N_{\mathrm{jets}} =2\,. 
\label{eq:charged_multiplicity}
\end{equation}

The multiplicity is sensitive to the development of the parton shower and to the subsequent hadronisation process. In QCD, its energy dependence can be described within the modified leading-logarithmic approximation for the basic calculation of the particle collision, alongside the local parton hadron duality hypothesis, which states how the momentum distribution from the final-state charged hadrons will closely mimic the distribution from the parton shower. Together, they will relate the multiplicity of the final state hadrons to the perturbative calculations from the partonic stage.
The parton multiplicity depends on the evolution scale $Q$ and on the running strong coupling $\alpha_s(Q)$, as well as on the colour factors and the number of active quark flavours.
\begin{figure}[!ht]
    \centering
    \includegraphics[width=0.96\linewidth]{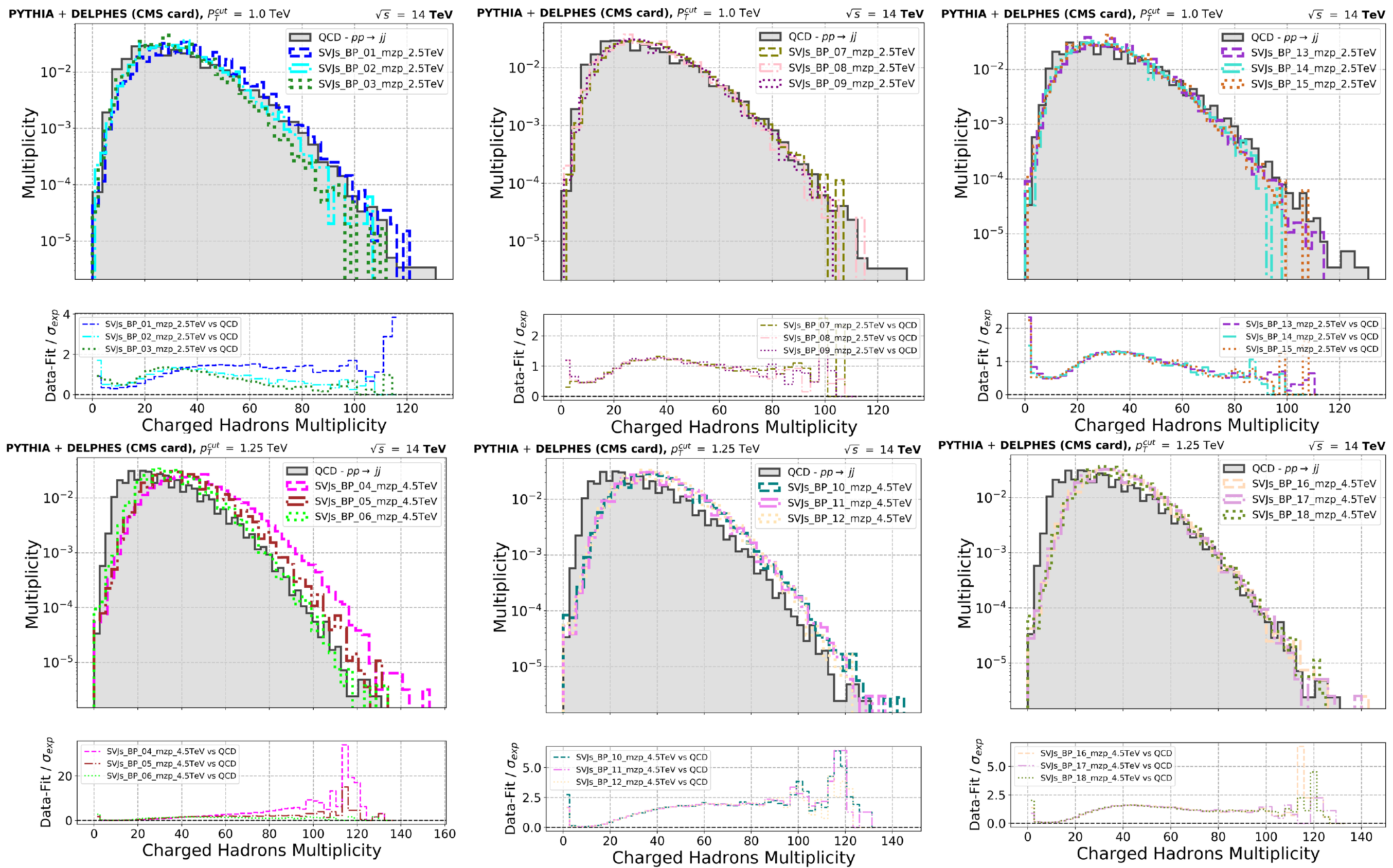}
    \caption{Charged hadrons multiplicity  for the QCD background and the full
set of SVJ BPs.}
    \label{fig:Hadrons_multi}
\end{figure}
At sufficiently high energies, the multiplicity is therefore controlled primarily by the perturbative evolution of the parton shower, while the overall normalisation contains information about the transition from partons to hadrons. This provides a useful connection between the perturbative shower and the observable charged particle multiplicity. In the dark sector, a similar picture applies when the hidden gauge group is QCD-like. The dark partons undergo a shower governed by the corresponding dark sector coupling and subsequently hadronise at the confinement scale $\Ldark$. The resulting dark hadrons can be either stable and invisible or unstable and decay back to visible SM particles. Consequently, the observed charged hadron multiplicity depends on both the development of the dark shower and the subsequent decay chain of the dark hadrons, given by the increase of the mediator mass $m_{Z^\prime}$ leading to higher observed charged hadron multiplicity.

The multiplicity provides a complementary probe of the dark-sector dynamics. Parameters that modify the dark parton shower can change the number and kinematic distribution of dark partons produced before hadronisation. The confinement scale $\Ldark$ subsequently determines the transition to the non-perturbative regime and can therefore affect the resulting dark-hadron spectrum. Since only the visible dark hadrons contribute to the reconstructed charged particle multiplicity, the final observable also depends on their decay modes and on the fraction of stable invisible states. Such behaviour is clearly illustrated in figure \ref{fig:Hadrons_multi}.

The multiplicity is additionally correlated with the jet transverse momentum and radius. Increasing the available phase space for radiation can lead to a larger number of resolved constituents, while the jet radius determines how much of this radiation is captured by the reconstructed jet. These effects make the charged hadron multiplicity sensitive to both the hard process kinematics and the internal structure of the dark shower. 
\section{DL Analyses}
\label{sec:4}
In this section we investigate the ability to distinguish between SM jets and SVJs using the variables we have discussed in the previous section.  
We compare five Neural Networks (NNs) trained on the data samples simulated from the BPs listed in Table \ref{tab:SVJs_event_gen_configs} to perform the binary classification task of distinguishing between QCD jets and SV jets. The first is a ViT acting on the two dimensional Lund jet  plane image, which retains the spatial organisation of the radiation but not the global event kinematics. The second is the JetLOV architecture of \cite{Diaz:2023otq}, which acts directly on the C/A reclustered tree of the jet and learns its node features representing the LJP from the constituent four momenta. The ViT and JetLOV architectures extract relevant substructure representations for binary classification task.   
The third is an MLP acting on a compact vector of coarse-grained kinematic and jet substructure observables.  The fourth combines the image using the ViT and feature representations with the MLP in a single fusion network. Another fusion architecture is the fifth network we use; it combines the MLP with the JetLOV network.  The fusion methods explore whether the constituent feature extraction methods yield  information that are redundant or complementary.

All five models are trained on a common balanced dataset, with identical event selection, identical training and validation splits, and the same optimisation and early stopping protocol, so that differences in performance can be attributed to the representation and architecture rather than to the training configuration. Their performance is quantified on a held-out test sample through the Receiver Operating Characteristic (ROC) curves and we report the corresponding Area Under the Curve (AUC). In addition, all models are evaluated on simulated data from three BPs that are generated at an intermediate mediator mass of $3.5$ TeV (see table~\ref{tab:SVJs_event_gen_configs}) that were not used at any stage of the training.  This provides an independent test of whether the classifiers generalise across the dark sector parameter space rather than learning features specific to the training BPs.

\subsection{Events Generation and Data Preprocessing}
\label{sec:4.1}

In this work, we generate 18 signal BPs for SVJs and one QCD background sample for training and testing, and another 3 for testing transportability of performance (see table~\ref{tab:SVJs_event_gen_configs}).
Both SVJ and QCD events are generated using \textsc{PYTHIA8.3.17} \cite{Bierlich:2022pfr}, with QCD events generated as 
$pp \rightarrow jj$.
The generated events are passed to \textsc{DELPHES3.5.1} \cite{deFavereau:2013fsa} for detector simulation using the CMS detector card. Jets are reconstructed with \textsc{FastJet3.5.1} \cite{Cacciari:2011ma} using the anti-$k_T$ algorithm \cite{Cacciari:2008gp} with $R=0.8$. This radius provides a compromise between retaining the visible radiation from the dark shower and hadronisation and maintaining a well defined di-jet topology. A larger radius would capture more radiation but increase the overlap between neighbouring jets.

For the DL analysis, we apply the event selection summarised in table~\ref{tab:SVJs_cuts_ML}. We require at least two large-$R$ jets, with the leading jet satisfying $p_T>1.0~\mathrm{TeV}$ for $m_{Z^\prime}=2.5~\mathrm{TeV}$ and $p_T>1.25~\mathrm{TeV}$ for $m_{Z^\prime}=4.5~\mathrm{TeV}$. Both leading jets are required to satisfy $|\eta|<2.1$. We further impose requirements on the missing transverse momentum and the angular separation of the two leading jets. These selections retain the characteristic SVJ topology while providing sufficient statistics for the subsequent DL analysis.

\begin{table}[!ht]
    \centering
    \renewcommand{\arraystretch}{1.1}
        \begin{tabular}{|l|c|}
            \hline
            \textbf{Cut} & \textbf{Requirement} \\
            \hline
            Large-$R$ jets multiplicity & $\geq 2$ jets \\
            \hline
            Large-$R$ jet $p_{T}$ (Leading Jet at $Z^{\prime}:m0~=~2.5~\text{TeV}$) & $p_{T} > 1.0~\text{TeV}$ \\
            \hline
            Large-$R$ jet $p_{T}$ (Leading Jet at $Z^{\prime}:m0~=~4.5~\text{TeV}$) & $p_{T} > 1.25~\text{TeV}$ \\
            \hline
            Large-$R$ jet pseudorapidity & $|\eta| < 2.1$ \\
            \hline
            Missing transverse energy & $E_{T}^{\text{miss}} > 200~\text{GeV}$ \\
            \hline
            Azimuthal separation & $\Delta\phi(j_1, j_2) > 0.8$ \\
            \hline
            Rapidity separation & $|\Delta y(j_1, j_2)| < 2.8$ \\
            \hline
        \end{tabular}
    \caption{Event selection cuts for the ML analyses.}
    \label{tab:SVJs_cuts_ML}
\end{table}

The selected events are subsequently used to construct the combined dataset. To account for the different production rates and selection efficiencies of the BPs, we determine the selection efficiency for each signal sample as
\begin{equation}
\epsilon =
\frac{N_{\rm pass}}{N_{\rm total}},
\end{equation}
where $N_{\rm pass}$ denotes the number of events passing the analysis requirements and $N_{\rm total}$ is the total number of generated events. The corresponding effective cross section is then given by
\begin{equation}
\sigma_{\rm eff} = \sigma_{\rm prod} \ \epsilon,
\end{equation}
where $\sigma_{\rm prod}$ is the production cross section before the analysis selection. The effective cross sections are used to assign relative weights to the different SVJ BPs when constructing the combined dataset for DL training.

For the DL training, we construct balanced datasets containing 300,000 SVJ signal events and 300,000 QCD background events. The signal sample is obtained by combining events from the 18 BPs according to their respective effective cross sections. Each event is represented by three complementary inputs, as follows.
\begin{itemize}

\item The C/A reclustered tree dressed with the 4-momentum information at each node.

\item LJP images, constructed by binning the SVJ constituents on a discretised $50\times50$ grid. The pixel intensity is given by the corresponding LJP density.

\item A 15-dimensional vector of kinematic and jet substructure observables.
\end{itemize}

The feature vector contains the missing transverse momentum and its azimuthal angle, that we call global variables.  Local, or jet level features, are those that aggregate kinematic information from within each of the two leading jets that we dub final state, and those that extract correlations between jet constituents, or jet substructure variables. The final state variables include $p_T$, $\eta$, and $\phi$ of the two leading fat jets, the di-jet invariant mass $M_{\rm jj}$, the constituent hadron multiplicity, and the azimuthal separation between the two leading jets.  The substructure variables are  the three angularities for $\alpha = 0.1, 0.8, 1.5$, and the two-point energy correlation function. These observables probe both the global event kinematics and the internal radiation pattern of the jets, and feed into a MLP classifier, to be described below.

\begin{table}[htbp]
\centering
    \begin{tabular}{ccc}
    \toprule
            \multirow{2}{*}{\textbf{Global}} & \multicolumn{2}{c}{\textbf{Local}} \\
            \cmidrule(lr){2-3}
             & \textbf{Jet-level} & \textbf{Sub-jet level} \\
            \midrule
             $\text{MET}$ in eq(\ref{eq: MET})& $p_T(j_1),\,p_T(j_2)$  \\
             $\Delta\phi(\mathrm{MET},j_{\mathrm{lead}})$ in eq(\ref{eq: diff_azimuthal}) & $\phi(j_1),\,\phi(j_2)$ & $\tau(0.1)$ in eq(\ref{eq: angularity})\\
             & $\eta(j_1),\,\eta(j_2)$ & $\tau(0.8)$ in eq(\ref{eq: angularity})\\
             & $M_{\rm jj}$  in eq(\ref{Eq:di-jet_mass}), $\Delta\phi(j_1,j_2)$ & $\tau(1.5)$ in eq(\ref{eq: angularity})\\
             & $\langle n_{\mathrm{ch}}\rangle$ in eq(\ref{eq:charged_multiplicity}) & $e_2$--EEC in eq(\ref{eq:e2_eec})\\
    \bottomrule
    \end{tabular}
\caption{The 15 high-level MLP input observables employed, categorized in three classes, for the global-event level variables, and the local jet level variables
            (this last category also includes the final-state and the jet substructure metric). The MLP was only employed for the JetLOV and the ViT analyses.}
\label{tab: MLP_input_observables}
\end{table}

The jet constituent trees, LJP images and kinematic features are used independently to train the respective networks. For the fusion models, the kinematic features are restricted to the same event subset as the images, with identical event indices used for both representations. Thus, each image-feature pair corresponds to the same event. 

All kinematic and jet substructure inputs are standardised using statistics computed from the training sample,
\begin{equation}
x_{\mathrm{std}} = \frac{x-\mu}{\sigma}\,.
\label{eq:std}
\end{equation}
The normalisation parameters are stored as non-trainable buffers and are therefore included in the model checkpoint.
To evaluate the ability of the DL models to interpolate across the model parameter space, we introduce three additional BPs that are not included in the training samples. These points are chosen within the parameter ranges covered by the 18 training BPs and are listed as the last three BPs in table~\ref{tab:SVJs_event_gen_configs}. In particular, all three testing points are generated with an intermediate mediator mass, $ \mZp =3.5~\mathrm{TeV}$, 
which lies between the two mediator masses used for training, $m_{Z^\prime}=2.5$ and $4.5~\mathrm{TeV}$. The remaining dark sector parameters are varied across the testing points to probe different regions of the parameter space. Since these configurations are not seen during training, they provide an independent test of whether the trained models can identify SVJs and retain their performance for parameter values that were not explicitly included in the training dataset.
\subsection{ViT}
\label{sec:4.2}
In contrast to conventional jet images, the LJP directly encodes information on the momentum and angular structure of the constituents, making it sensitive to differences in the underlying shower and hadronisation dynamics. In particular, the dark shower can produce a broader and more diffuse radiation pattern through the production and subsequent decay of dark hadrons. The resulting correlations can extend over different angular scales and need not be confined to local regions of the image. In our case, we choose the local region within the ranges of the LJP kinematical axes, where the values were given for each variables as $\kappa~\in~\{-2,6\}$ and $\lambda~\in~\{0.0,7.5\}$, for the preprocessing, the binning from each of the distributions should be comparable to the resolution.

We therefore employ a ViT network to analyse the LJP images. The ViT is well suited to this task because self-attention does not impose a fixed locality prior. Each image patch can interact directly with all other patches, allowing the network to identify correlations between radiation at widely separated locations in the LJP. This is useful for distinguishing SVJs from QCD jets, where the relevant information is not necessarily associated with a single localised feature but can be distributed over the full radiation pattern. The self-attention mechanism can also correlate radiation patterns across different angular scales, providing sensitivity to both the collimated radiation associated with the perturbative shower and the wider angle radiation produced by dark hadron decays.

Each $50\times50$ image is divided into 25 non-overlapping $10\times10$ patches. Each patch is flattened and mapped to a 48-dimensional embedding,
\begin{equation}
e_i = W_p \ x_i + b_p,  \qquad i=1,\ldots,25, 
\label{eq:patch}
\end{equation}
where $x_i$ denotes the $i$-th image patch. A learnable class token, $x_{\mathrm{class}}\in\mathbb{R}^{48}$, is appended to the sequence of patch embeddings, and a learnable positional embedding, $E_{\mathrm{pos}}$,  is added to retain information about the spatial location of each patch:
\begin{equation}
z_0 = \left[ x_{\mathrm{class}}; e_1; \ldots; e_{25} \right] + E_{\mathrm{pos}}, \qquad E_{\mathrm{pos}}\in\mathbb{R}^{26\times48}. 
\label{eq:tokens}
\end{equation}
The positional embedding is required because self-attention is permutation invariant and therefore does not preserve the ordering of the input tokens. It enables the network to distinguish between radiation patterns occurring at different locations in the LJP, for example between radiation concentrated near the jet core and radiation at larger angular separations.

The resulting sequence is processed by six Transformer encoder blocks with pre-layer normalisation. Each block first applies four multi-head self-attention, allowing the representation of each patch to incorporate information from all other patches, followed by a feed forward network with dimensions $64\to128\to 64$. Residual connections are used around both sublayers, to preserve the information from the previous layer while allowing the attention and feed-forward sublayers to learn increasingly informative representations of the LJP radiation pattern.

For each attention head, the attention weights are given by
\begin{equation}
\mathrm{Attention}(Q,K,V)   =   \mathrm{softmax} \left(   \frac{QK^T}{\sqrt{d_k}}   \right)V \,,
  \label{eq:attention}
\end{equation}
where we follow the standard conventions for the attention mechanism, with $Q$, $K$, and $V$ denoting the query, key, and value representations, respectively. This operation enables the model to assign different weights to different regions of the LJP and to learn which combinations of radiation patterns are most relevant for classification. The feed-forward network subsequently provides a nonlinear transformation of the aggregated information. We use the Gaussian error linear unit (GELU) activation, and after the final encoder block, the class token representation is passed through a LayerNorm to obtain a 48-dimensional image embedding, $h_{\mathrm{img}}$. For the standalone image classifier, this embedding is mapped to a single logit,
\begin{equation}
\mathrm{logit} = W_h\  h_{\mathrm{img}}+b_h, 
\end{equation}
which is converted to a classification probability using the sigmoid function which is used to classify the SVJs from the QCD events.

The use of self-attention allows the network to retain correlations between radiation at different angular scales, which can arise from the interplay between the dark parton shower, dark hadron production and subsequent decays. The ViT therefore provides a suitable architecture for learning the non-local features that distinguish the internal structure of SVJs from that of QCD jets. The resulting image classifier is used as a standalone model and, in the fusion architecture discussed below, its learned image representation is combined with the complementary kinematic and jet substructure information.

\subsection{JetLOV Network}
\label{sec:4.3}

The JetLOV architecture~\cite{Diaz:2023otq} combines a trainable feature extraction network, RegNet \cite{Radosavovic2020regnet}, with the LundNet architecture \cite{Dreyer:2020brq}. RegNet learns an analogous representation to that used in the LundNet architecture directly from the four momenta of the nodes in the binary clustering tree $\mathcal{T}$ without having to precompute the conventional Lund variables. This representation is then passed to LundNet-like architecture for jet classification.  This allows the network to learn features relevant for discrimination without imposing the conventional Lund variables as a fixed representation.

From each split indexed by $k$ in a C-A clustering tree, a parent node $k$ is associated with two daughter nodes $i$ and $j$, $p_k\rightarrow p_i + p_j$ and provided as input to RegNet: 
\begin{equation}
\left(p_i,p_j\right)_k
\xrightarrow{\mathrm{RegNet}}
\mathbf{v}_k \in \mathbb{R}^d.
\end{equation}
In \cite{Diaz:2023otq}, $d$ was chosen to be $5$ and initialised to recover the $5$-dimensional LundNet variables,
\begin{equation}
\left(p_i,p_j\right)_k
\xrightarrow{\mathrm{RegNet}}
\mathbf{v}^{(0)}_k = \left( \ln k_t , \ln\Delta , \ln z , \ln m , \psi \right)_k. 
\label{eq:JetLOV_mapping}
\end{equation}
The five components correspond to the variables conventionally used as inputs to LundNet. RegNet is implemented as a MLP with five output branches, with each branch learning one component of the representation. It therefore provides a learnable mapping from the local kinematics of each splitting to a compact set of node  features.

The training is performed in two stages. First, RegNet is trained as a regression network to reproduce the conventional Lund variables calculated from the daughter four momenta. This pre-training provides an initial representation compatible with the subsequent LundNet input. The pre-trained RegNet is then connected to a separately pre-trained LundNet, replacing the analytically constructed Lund variables. 
The complete JetLOV mapping can be summarised as
\begin{equation}
\left\{ (p_i,p_j)_k \right\}_{k\in\mathcal{T}} \xrightarrow{\mathrm{RegNet}} \left\{ \mathbf{v}_k
\right\}_{k\in\mathcal{T}} \xrightarrow{\mathrm{LundNet}} P(y|\mathcal{T}), 
\end{equation}
where $\mathcal{T}$ denotes the set of nodes in the jet clustering tree and $P(y|\mathcal{T})$ is the probability assigned to the jet class. During the regression stage, $\mathbf{v}_k$ is constrained to reproduce the conventional Lund variables $\mathbf{v}^{(0)}_k$. This constraint is removed during the subsequent classification stage, allowing RegNet and LundNet to be optimised jointly. The learned representation can therefore depart from the conventional Lund variables, and can also be of different dimension $d$, if an alternative representation provides more discriminating information.

In  this work, JetLOV provides a complementary representation of the SVJ structure to the LJP image used by the ViT. While the ViT learns spatial correlations in the two dimensional LJP, JetLOV operates directly on the hierarchical structure of the jet and learns node  features from the underlying four momenta. The JetLOV classifier is therefore used as an alternative summary of the dark shower and hadronisation information encoded in the jet substructure.

\subsection{MLP}
\label{sec:4.4}
The third classifier operates exclusively on the high level observables introduced in previously and provides a baseline against which the image-based and tree-based networks can be assessed. Whereas the ViT and JetLOV
architectures are designed to extract information from a structured representation of the jet, the MLP  receives a fixed length vector of physically motivated quantities and therefore quantifies the discrimination that is already available from the global event kinematics and from a small set of substructure observables, without any learned representation of the radiation pattern.
The input is the $15$ dimensional feature vector $x\in\mathbb{R}^{15}$ described in described in table \ref{tab: MLP_input_observables}.
The standardisation of Eq.~\eqref{eq:std} is applied as the first operation of the network rather than as an external preprocessing step, with the training-sample mean $\mu$ and standard deviation $\sigma$ registered as non-trainable buffers, so that the normalisation is carried by the model checkpoint and cannot be inconsistently applied at inference time 
The network consists of three fully connected blocks of decreasing width, $256\rightarrow 128 \rightarrow 64$. Each block applies an affine transformation followed by batch normalisation, a rectified linear unit ($\mathrm{ReLU}$) and dropout,
\begin{equation}
  h^{(l)} = \mathcal{D}_{p}\Bigl(
  \mathrm{ReLU}\bigl(
  \mathrm{BN}\bigl( W^{(l)} h^{(l-1)} + b^{(l)} \bigr)
  \bigr)\Bigr),
  \qquad l = 1,2,3,
  \label{eq:mlp_block}
\end{equation}
where $\mathcal{D}_{p}$ denotes dropout with rate $p=0.1$ and $\mathrm{BN}$ is the batch normalisation function. Batch normalisation stabilises the optimisation given the widely differing scales of the input observables, which range from dimensionless angularities of order unity to transverse momenta of several TeV, while dropout acts as a regulariser and limits the extent to which the network can rely on any individual feature. The final block is followed by a single linear layer producing one logit,  which is interpreted as the probability that the event originates from an SVJ
signal. The network contains $12,865$ trainable parameters, several orders of magnitude fewer than the ViT and JetLOV models, so that any comparable performance obtained with this architecture indicates that the corresponding discriminating information is already present in the high level observables
themselves.

The MLP therefore serves as a useful reference for assessing the additional information captured by the more structured representations. Its input observables encode global kinematics and selected aspects of jet substructure, but do not retain the spatial or hierarchical organisation of the radiation pattern. Comparing its performance with that of the ViT and JetLOV networks consequently allows us to determine whether the discrimination of SVJs from QCD is driven primarily by a small set of high level observables or by more detailed information contained in the LJP and clustering tree representations.

\subsection{Fusion Networks}
\label{sec:4.5}

The three classifiers described above operate on different representations of the same event; but these representations are not equivalent. The LJP image retains the spatial organisation of the radiation pattern but discards the global event kinematics; the clustering tree retains the hierarchical ordering of the splittings and the underlying constituent four momenta, but likewise carries no information on the event as a whole; and the high level observables provide a compact description of the event kinematics together with a few angular moments of the radiation, but retain neither the spatial nor the hierarchical organisation of the jet.

We therefore construct two fusion networks. The first combines the LJP image with the high level observables, and the second combines the clustering tree with the same set of observables. Both follow the same template, so that the two can be compared directly with each other as well as with the corresponding standalone classifiers.

\begin{figure}[!ht]
\centering
\includegraphics[width=0.96\linewidth]{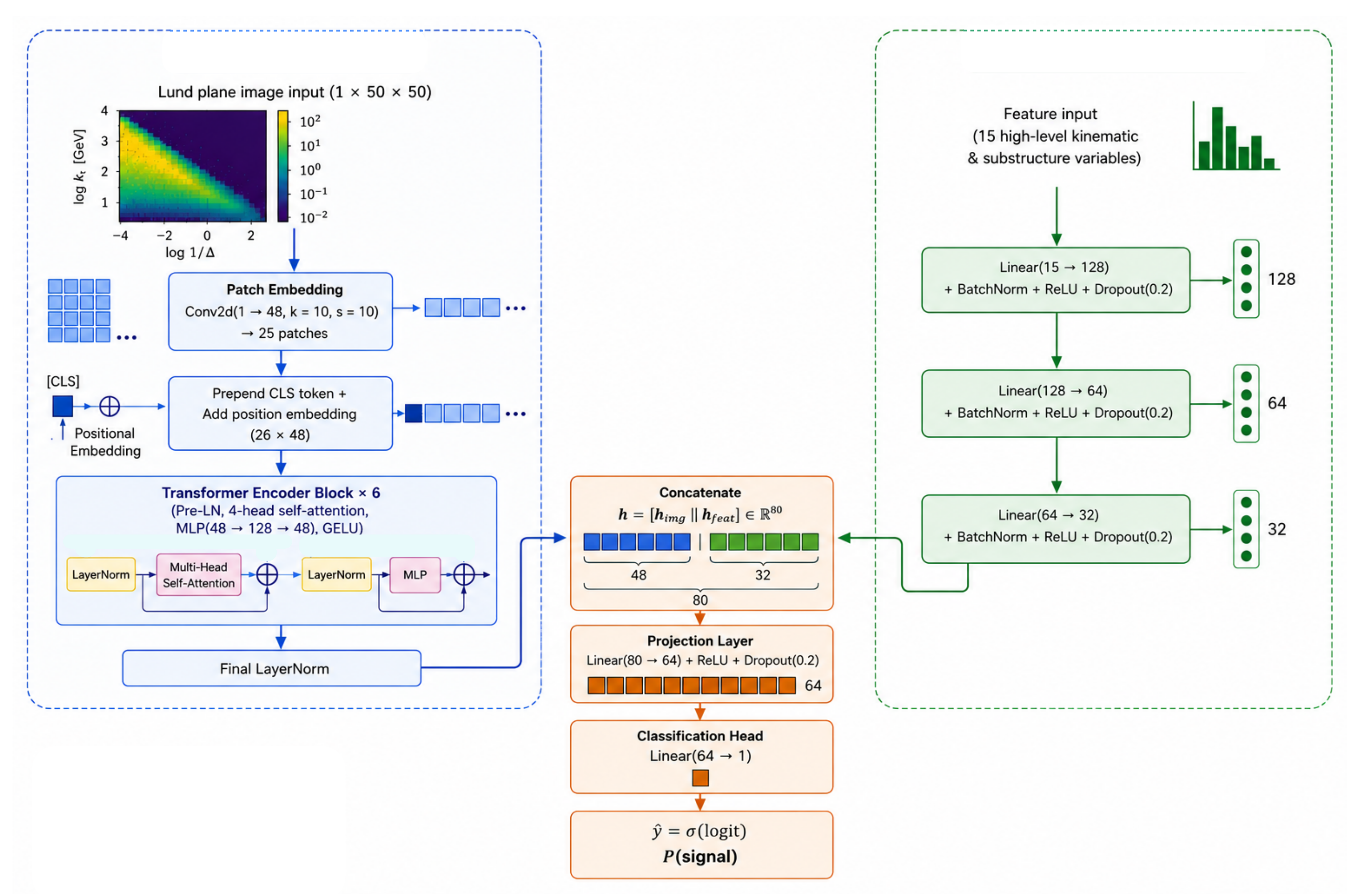}
\caption{Schematic architecture of the ViT and MLP  fusion network.}
\label{fig:fusion_network}
\end{figure}
In both cases the two branches act as encoders, with their classification heads
removed, and therefore produce learned embeddings rather than classification
probabilities. The feature branch is in both cases the MLP of
section~\ref{sec:4.4} with its final linear layer removed, so that its output is
the activation of the last hidden block,
\begin{equation}
h_{\mathrm{feat}} = \mathrm{MLP}(x) \in \mathbb{R}^{64},
\end{equation}
where $x$ is the $15$ dimensional feature vector. Writing $h_{\mathrm{jet}}$ for the embedding produced by either the ViT or JetLOV branch, the two are concatenated and passed through a projection layer, followed by a single linear layer producing the classification logit. Each branch retains its own input normalisation, stored
as non-trainable buffers so that it is carried by the model checkpoint; the high level features are standardised per component according to Eq.~\eqref{eq:std}, while the structured branch normalises its own input as described below.

Combining the embeddings before the final classification layer, rather than averaging two independent classification scores, allows the network to learn nonlinear correlations between the internal structure of the jet and the global event kinematics. The joint optimisation also allows each encoder to adapt its representation to the information provided by the complementary branch, as demonstrated in \cite{Hammad:2023sbd,Hammad:2024hhm}.

\paragraph{ViT and MLP Fusion} 
The image branch is the ViT encoder of section~\ref{sec:4.2}. The $50\times50$ LJP image is divided into $25$ non-overlapping $10\times10$ patches, which are linearly embedded, augmented with a class token and a learnable positional embedding, and processed by six Transformer encoder blocks with four attention heads each. The class token representation after the final layer normalisation gives the image embedding, $h_{\mathrm{img}} \in \mathbb{R}^{128}$, which is concatenated with MLP output branch. The images are normalised using a single global mean and standard deviation computed over the training partition.

This model contains $637,345$ trainable parameters, of which $614,144$ belong to the image branch, $12,832$ to the feature branch, and $10,369$ to the projection layer and the classification head. Both encoders are initialised randomly and optimised jointly, with the gradients of the classification loss propagated through both branches. The full network is depicted in figure \ref{fig:fusion_network}.

\paragraph{JetLOV and MLP Fusion}
The second fusion network replaces the image branch by the JetLOV encoder of section~\ref{sec:4.3}, so that the structured representation is the Lund declustering tree rather than its two dimensional projection. For each node $k$ of the tree, RegNet maps the concatenated four momenta of the two daughters to the five dimensional representation of Eq.~\eqref{eq:JetLOV_mapping}. The resulting node features are propagated through the LundNet graph network, which consists of six EdgeConv blocks with channel pairs $(32,32)$, $(32,32)$, $(64,64)$, $(64,64)$, $(128,128)$ and $(128,128)$. The outputs of the six blocks are concatenated, averaged over the nodes of the jet and mapped to the jet embedding, $h_{\mathrm{jet}} \in \mathbb{R}^{256}$, which is concatenated with MLP output branch. The node input features are standardised componentwise using statistics computed over the nodes of the training partition.

Unlike the image branch, the JetLOV branch is not initialised randomly. As discussed in section~\ref{sec:4.3}, RegNet is first pre-trained as a regression network against the analytically computed Lund variables, since a randomly initialised RegNet is not able to reach a competitive minimum in the subsequent classification stage. The pre-trained weights are then loaded into the fusion model, then the learned node features are free to depart from the conventional Lundvariables during the joint optimisation.

This model contains $324,314$ trainable parameters, of which $58,521$ belong to RegNet, $234,400$ to the LundNet graph network, $12,832$ to the feature branch, and $18,561$ to the projection layer and the classification head. It is therefore roughly half the size of the image-based fusion model, despite
acting on a less compressed representation of the jet.
\\

Because the two fusion models share the same feature branch, the same fusion  layer and the same optimisation, the comparison between them isolates the contribution of the structured representation. Their comparison with the standalone ViT, JetLOV and MLP classifiers establishes whether the information contained in the radiation pattern is largely redundant with the high level observables or provides additional discrimination, and second, whether the two dimensional LJP projection or the full clustering tree provides the more effective description of the dark shower for this purpose.

\section{Results and Discussion}
\label{sec:5}

The training configurations used for the five classifiers are summarised in table \ref{tab:trainconfig}. We assess the performance of the five classifiers on a common held-out test sample, as shown in table~\ref{tab: Results_ML_analysis}. [1--18]BPs are used for training and early stopping, while the test sample was based on [19--21]BPs, which are statistically independent from the training samples. We evaluate each classifier using the accuracy, the area under the receiver operating characteristic curve (ROC), and the precision, recall and $F_1$ score for the signal class, with the latter quantities evaluated at a decision threshold of $0.5$. 

Before comparing the classification performance, we briefly discuss the training behaviour of the different architectures.  All models exhibit stable convergence within the considered training ranges. The MLP reaches its minimum validation loss after eight epochs, corresponding to approximately one minute of training. The ViT, ViT+MLP and JetLOV+MLP models reach their respective low loss configurations after 30, 43 and 13 epochs, while the JetLOV model reaches its best validation loss after seven epochs. The substantially longer training time of the JetLOV models is primarily associated with the computational cost of processing the graph representations. The attention architectures in ViT models also require more epochs and computational resources than the MLP, reflecting their larger model complexity and the associated optimisation cost.

For all classifiers, the training and validation losses remain comparable around the best epoch, with no pronounced divergence indicative of overfitting. The early stopping procedure terminates training before a significant separation between the two losses develops. Within the explored training ranges, the reported results are consequently not limited by an apparent overfitting of the training sample. Nevertheless, it should be noted that the minimum validation loss does not provide a direct measure of the expected test set ranking across architectures. In particular, the classifiers differ in their representational capacity, optimisation dynamics and regularisation, such that a lower validation loss for one architecture need not imply a higher test set performance relative to another. 
\begin{table}[!ht]
\centering
\small
\setlength{\tabcolsep}{4.5pt}
\begin{tabular}{lccccc}
\toprule
Hyperparameter & ViT & MLP & JetLOV & ViT+MLP & JetLOV+MLP \\
\midrule
Optimizer                 & Adam & Adam & Adam & Adam & Adam \\
Learning rate             & $7\times10^{-4}$ & $1\times10^{-3}$ & $1\times10^{-3}$ & $5\times10^{-4}$ & $5\times10^{-4}$ \\
Adam ($\beta_1,\beta_2$)  & 0.9, 0.999 & 0.9, 0.999 & 0.9, 0.999 & 0.9, 0.999 & 0.9, 0.999 \\
Batch size                & 128 & 256 & 256 & 128 & 128 \\
Dropout                   & 0.1 & 0.2 & 0.1 & 0.1 / 0.2 & 0.1 / 0.2 \\
Weight decay              & 0 & 0 & 0 & 0 & 0 \\
Max epochs                & 60 & 100 & 15 & 60 & 60 \\
Early-stop patience       & 5 epochs & 5 epochs & 5 epochs & 5 epochs & 5 epochs \\
Early-stop min.\ $\Delta$ & $10^{-4}$ & $10^{-4}$ & $10^{-4}$ & $10^{-4}$ & $10^{-4}$ \\
Random seed               & 42 & 42 & 42 & 42 & 42 \\
\midrule
Epochs trained (max) & 35 (60)   & 11 (100) & 15 (20)  & 48 (60)   & 18 (69) \\
Best epoch           & 30        & 8        & 7        & 43        & 13\\
Best val.\ loss      & 0.3361    & 0.2356   & 0.2518   & 0.1567    & 0.1504 \\
Avg.\ epoch time     & 75.2\,s   & 5.3\,s   & 516.7\,s & 63.8\,s  & 335\,s \\
Total train time     & 43.8\,min & 1.0\,min & 1.01\,hrs & 51.0\,min & 1.68\,hrs \\
\bottomrule
\end{tabular}
\caption{Training hyperparameters for the five classifiers.}
\label{tab:trainconfig}
\end{table}

The resulting classification performance is shown in table~\ref{tab:results}. Among the classifiers based solely on the internal structure of a single jet, JetLOV substantially outperforms the ViT, reaching an accuracy of $0.9003$ and an AUC of $0.9572$, compared with $0.8629$ and $0.9293$ for the ViT. This corresponds to a reduction in the misclassification rate from $13.7\%$ to $10.0\%$. Both classifiers are constructed from the same C/A  declustering sequence of the leading large-$R$ jet, so the difference is attributable to the representation of the information rather than to the underlying physical input.

\begin{table}[!ht]
\centering
\begin{tabular}{lccccc}
\toprule
Model & Accuracy & ROC AUC & Precision (sig.) & Recall (sig.) & $F_1$ (sig.) \\
\midrule
ViT                             & 0.8629 & 0.9293 & 0.8731 & 0.8488 & 0.8608 \\
JetLOV                          & 0.9003 & 0.9572 & 0.9222 & 0.8735 & 0.8972 \\
MLP (All)                       & 0.9054 & 0.9653 & 0.9156 & 0.8930 & 0.9041 \\
MLP (Substructure)             & 0.8570 & 0.8861 & 0.8542 & 0.8661 & 0.8480 \\
MLP (Jet-level)                     & 0.8676 & 0.8920 & 0.8630 & 0.8725 & 0.8627 \\
ViT+MLP                         & 0.9389 & 0.9846 & 0.9405 & 0.9370 & 0.9387 \\
JetLOV+MLP                      & \textbf{0.9428} & \textbf{0.9872} & \textbf{0.9560} & \textbf{0.9366} & \textbf{0.9462} \\
\bottomrule
\end{tabular}
\caption{Classification performance of the five networks, with the best performing result highlighted in bold.}\label{tab:results}
\label{tab: Results_ML_analysis}
\end{table}

There is also a statistical advantage to the tree-based representation. The LJP image contains $2500$ pixels for typically only $\mathcal{O}(10)$ resolved splittings and is consequently highly sparse. The image encoder must therefore learn to process a large number of empty pixels, whereas the graph representation retains only the populated nodes. Consistent with this observation, JetLOV achieves its performance with roughly one third of the parameters of the image encoder.

The MLP, which uses 15 high level observables, performs slightly better than JetLOV, with an accuracy of $0.9054$ and an AUC of $0.9653$. This comparison  should be interpreted with some care. Unlike the jet level classifiers, the MLP has access to global event information, including the missing transverse momentum, the transverse momenta of the two leading jets and the di-jet invariant mass. It therefore directly probes the characteristic event level signature of semivisible production, namely genuine missing transverse momentum correlated with the visible jets. 
Another two MLP networks were added where we took each of the kinematical analyses separately, one for jet substructure observables and the other for jet-level observables, based on their classification in table \ref{tab: MLP_input_observables}. The substructure-only case is the worst performing, with an accuracy of only $0.8570$, while the jet-level case is only slightly better than the ViT with $0.8676$. The fact that the MLP with all analyses performs significantly better than either of these shows that there is complementary information in the different types of analysis.
Despite this difference, the comparable performance of JetLOV and the MLP is notable. A classifier based  on the clustering tree of one jet recovers most of the discrimination obtained from global event kinematics, including the missing energy information. This demonstrates that the visible radiation pattern of a SVJ provides a substantial discriminating handle in its own right. This complementarity is particularly relevant in regions with a small invisible fraction, where the missing energy signature becomes less pronounced while modifications of the jet subjet structure can remain visible.

The combination of complementary representations provides the strongest discrimination. The ViT+MLP fusion network reaches an accuracy of $0.9389$ and an AUC of $0.9846$, improving on both individual branches. The misclassification rate decreases from $9.5\%$ for the MLP and $13.7\%$ for the ViT to $6.1\%$, corresponding to a relative reduction of about $35\%$ with respect to the performing standalone classifiers. Since the two branches are evaluated on the same events and trained jointly, this improvement demonstrates that the LJP representation contains information that is not captured by the high level observables.

The complementarity of the two representations is also physically intuitive. The constituent multiplicity, angularities  and the two-point correlator  correspond to particular weighted moments of the radiation pattern. While these observables provide informative projections of the jet structure, they do not constitute a complete description of the underlying radiation pattern. The gain from the fusion network therefore quantifies the additional information retained by the LJP representation. Conversely, the improvement of the fusion model over the standalone ViT shows that global event kinematics cannot be reconstructed from the jet image, which is constructed from jet constituents and is barely sensitive to global event information.

The JetLOV+MLP fusion network provides a further improvement, achieving the best overall performance with an accuracy of $0.9428$, an AUC of $0.9872$. Its misclassification rate of $5.7\%$ is lower than that of both the standalone MLP ($9.5\%$) and the ViT+MLP fusion network ($6.1\%$), corresponding to a relative reduction of approximately $40\%$ and $6\%$, respectively. The improvement over the standalone JetLOV is similarly substantial, demonstrating that the high level event observables provide complementary information to the hierarchical jet representation. In particular, the JetLOV branch captures correlations encoded in the declustering tree, while the MLP incorporates global event kinematics and missing transverse momentum. Their combination therefore exploits both the modified internal radiation pattern of the semivisible jet and the event level imbalance induced by the invisible component. The superior performance of JetLOV+MLP over ViT+MLP further indicates that the tree representation provides a more effective complement to the global observables than the corresponding LJP image.

Additional  classification performance is summarised by the row-normalised confusion matrices in table~\ref{tab:confusion}.
\begin{table}[!ht]
\centering
\begin{tabular}{lcccc}
\toprule
Model & True neg.\ (TN) & False pos.\ (FP) & False neg.\ (FN) & True pos.\ (TP) \\
\midrule
ViT       & 87.68\% & 12.32\% & 15.12\% & 84.88\% \\
JetLOV    & 91.19\% & 8.81\% & 15.06\% & 84.94\% \\
MLP       & 91.78\% &  8.22\% & 10.70\% & 89.30\% \\
ViT+MLP   & 94.08\% &  5.92\% &  6.30\% & 93.70\% \\
JetLOV+MLP & 94.99\% &  5.01\% &  6.34\% & 93.66\% \\
\bottomrule
\end{tabular}
\caption{Row-normalised confusion matrices on the test set, with background and signal corresponding to the negative and positive classes, respectively.}
\label{tab:confusion}
\end{table}
For collider searches, performance at a fixed decision threshold does not fully capture the behaviour of analyses in which the background is dominated, where stringent background suppression is often required. The false positive rate in table~\ref{tab:confusion} corresponds to a single working point, whereas the ROC  curves allow the signal efficiency to be evaluated across a continuous range of background rejection. Table~\ref{tab:roc} therefore reports the signal efficiency at several fixed background rejection values. The relative advantage of the fusion networks becomes increasingly pronounced as the background rejection is tightened. At $80\%$ background rejection, JetLOV+MLP improves the signal efficiency by about three percentage points relative to the MLP, while at $99\%$ rejection the improvement reaches fifteen percentage points, with $\epsilon_S=0.819$ for JetLOV+MLP compared with $0.643$ for the MLP and $0.508$ for the ViT.

\begin{table}[htbp]
\centering
\begin{tabular}{lcccc}
\toprule
Background rejection & ViT & MLP & ViT+MLP & JetLOV+MLP \\
\midrule
99\% & 0.508 & 0.643 & 0.794 & 0.819 \\
95\% & 0.728 & 0.845 & 0.927 & 0.937 \\
90\% & 0.823 & 0.908 & 0.964 & 0.969 \\
80\% & 0.896 & 0.954 & 0.985 & 0.988 \\
\bottomrule
\end{tabular}
\caption{Signal efficiency at different background rejection values.}
\label{tab:roc}
\end{table}

This improvement can be directly related to the expected sensitivity of SVJ identification searches. In the limit where the statistical significance is approximately given by
\begin{equation}
\frac{S}{\sqrt{B}}\propto\frac{\epsilon_S}{\sqrt{\epsilon_B}} \,,
\end{equation}
where $S,B$ are the number of SVJs and QCD events after optimizing the cuts on the network's output, while $\epsilon_s$ is the signal significance at fixed background rejection value, $\epsilon_B$.
At fixed background rejection, the ratio of signal efficiencies therefore directly measures the corresponding gain in the expected significance. At $99\%$ background rejection, the fusion network gives a factor of $1.27$ improvement relative to the MLP and a factor of $1.61$ relative to the ViT. At $80\%$ rejection, the corresponding gains are $1.04$ and $1.10$.

Several limitations should be kept in mind when interpreting these results. First, the high-level feature vector contains the di-jet invariant mass and the transverse momenta of the two leading jets, while the training sample combines BPs with $m_{Z^\prime}=2.5$ and $4.5~\mathrm{TeV}$ with a single QCD background sample. Part of the MLP discrimination may therefore arise from reconstructing the mediator resonance rather than from the dark shower dynamics themselves. Such information would not define a generic SVJs tagger and could lead to degraded performance for mediator masses outside the training range. The jet level classifiers are less susceptible to this effect, since neither the clustering tree nor the LJP image carries direct information about the mediator mass beyond that induced by the jet kinematics. The independent BPs at $m_{Z^\prime}=3.5~\mathrm{TeV}$ provide a direct test of this issue.

Second, all samples are generated with a single parton-shower and hadronisation model. The observables that drive the classification, particularly the constituent multiplicity and the density of resolved splittings at small $k_t$, are sensitive to the modelling of the non-perturbative regime in both QCD and the dark sector. The quoted performance should therefore be regarded as an estimate based on the chosen simulation setup rather than a direct projection of experimental sensitivity. Comparisons with independent shower and hadronisation models will be required to assess the robustness of the observed gains.

Finally, each architecture is trained only once, so the quoted results do not include uncertainties associated with random initialisation or minibatch ordering. The hyperparameters are also fixed independently for each architecture rather than optimised in a common procedure. Moreover, the reported performance is obtained for the combined signal sample. Since the BPs span different invisible fractions and dark hadronisation scales, a benchmark-by-benchmark study is required to determine where in parameter space the gain from combining the different representations is most pronounced.

\section{Conclusions}
\label{sec:6}

As we have shown, a comprehensive analysis of the various jet observables applied to SVJ has been produced. While some of these observables have been used in current CMS/ATLAS searches, none of the analyses have covered the full range of HV parameters with $Z^{\prime}$ variations. This includes most of the kinematic space that many jet observables could potentially cover. We then applied classification tasks to determine the discrimination performance between SM and HV signals for future collider searches. We also present how three specific HV parameters (e.g. dark quark mass, invisible fraction, and dark hadronisation) could assist in classifying signals from QCD to SVJ. The addition of new observables, such as the global MET, difference in azimuth, LJP, hadron multiplicity and 2-point EEC, enabled us to measure different stages of the HV process, such as ISR, soft and hard emissions, and MPI. No previous study has demonstrated this. Finally, although classification is not a new task in the search for these jets, previous studies have focused on one type of neural network. We present five new DL models, which provide a broader understanding of HV physics compared to QCD. Despite the detailed analysis of our results, further work is required to enable comparison between different BPs.

Overall, we have performed an study of SVJ identification in events produced through a heavy $Z^\prime$ mediator, using a set of $18$ BPs spanning two mediator masses and variations of the dark quark mass, invisible fraction, and dark hadronisation scale. Rather than relying on a single event representation, we compare classifiers based on three complementary descriptions of the same events: LJP images, hierarchical declustering trees, and a compact vector of global kinematic and substructure observables. We further consider two fusion networks that combine the structured jet representations with the high level observables. Since all models are trained on identical samples using a common selection and optimisation procedure, this setup provides a controlled comparison of the information captured by the different representations.

The comparison between the two jet level representations is equally informative. Both are constructed from the same declustering sequence, yet JetLOV reduces the misclassification rate from $13.7\%$ to $10.0\%$ relative to the image-based classifier, while using roughly one third of the number of parameters. The LJP image provides a binned representation of the emission density, with each splitting contributing independently to a pixel and the resulting representation being invariant under permutations of the splittings. In contrast, the tree representation preserves the ordering of the splittings and the branching relations between successive emissions. The observed performance gap therefore indicates that a non-negligible fraction of the discriminating information is encoded in these correlations rather than in the distribution of individual splittings alone. This is consistent with the expected dynamics of the dark shower, the invisible component removes energy at intermediate stages of the evolution, so that the visible remnant is characterised not only by a modified distribution of individual splitting scales, but also by their correlated sequence along the branching tree.

Combining the jet representations with global observables further improves the discrimination performance. The JetLOV+MLP fusion network achieves the best overall performance, with an accuracy of $0.9428$ and an AUC of $0.9872$, compared with $0.9389$ and $0.9846$ for the ViT+MLP network and the rest of the networks. The improvement over the standalone MLP shows that the detailed radiation pattern contains information beyond that captured by the angularities, two-point correlator, and constituent multiplicity. This additional information becomes particularly valuable at high background rejection, where the remaining QCD background events can fluctuate towards the large multiplicity and diffuse radiation patterns characteristic of semivisible jets. Although these fluctuations can make QCD jets resemble the signal in terms of low dimensional observables, their detailed splitting structure remains governed by ordinary soft-collinear QCD radiation. By resolving this structure through the hierarchical clustering tree, JetLOV can distinguish such background fluctuations more effectively. This explains why the benefit of the fusion network becomes most pronounced in the background dominated regime that is most relevant for the sensitivity of the search.

Overall, these results demonstrate that preserving the hierarchical structure of jet radiation and combining it with global event information provides a robust and complementary strategy for SVJ discrimination. While further studies are needed to assess the dependence on simulation modelling and training fluctuations, the results highlight the potential of structured and multimodal representations for improving the sensitivity of future SVJ searches.

\section*{Acknowledgments}

M.A.A.-B. acknowledges the IRIDISX High-Performance Cluster (HPC) at the University of Southampton for its contribution to the completion of this project. The project was partially funded by SECIHTI (Secretaría de Ciencia, Humanidades, Tecnología e Innovación), the Mexican government science secretariat. SM is supported in part through the NExT Institute and STFC Consolidated Grant ST/X000583/1.
The work of MHS was supported by UK STFC grant ST/X00077X/1, by a Southampton Theory, Astrophysics and Gravity (STAG) Research Centre visiting grant, and by a CERN Associateship.

\bibliographystyle{JHEP}
\bibliography{biblo}
\end{document}